\documentclass[preprint,showpacs ,aps]{revtex4-1}
\usepackage{amssymb,latexsym}
\usepackage{amsmath} 
\usepackage{graphicx} \usepackage{epsfig}
\usepackage{amssymb} 
\usepackage[utf8]{inputenc}
\usepackage{color} \usepackage{slashed} 
\newcommand{\ba}{\begin{align}} \newcommand{\ea}{\end{align}}
\newcommand{\beq}{\begin{eqnarray}}
\newcommand{\eeq}{\end{eqnarray}} \newcommand{\nneeq}{\nonumber
\end{eqnarray}}

\newcommand{\bfig}{\begin{figure}}
\newcommand{\efig}{\end{figure}}

\usepackage{feynmf}
\usepackage{tikz}
\usepackage[toc,title,page]{appendix}

\usepackage[compat=1.1.0]{tikz-feynman}
\begin{document}
%\title{\bf\large{Lightfront QED, Stueckelberg field and Infrared divergence}} 
\title{\bf\large{Equivalence of two component spinor mechanism and four component spinor mechanism in top quark pair production}} 
\author{\bf Malvika Deo$^{a,\$}$, Anuradha Misra $^{b,\dagger}$, Sharada Subramaniam$^{c,\$\$}$, Radhika Vinze
$^{d, \dagger\dagger,}\footnote[5]{Address after July 12, 2024: Indian Institute of Science Education and Research Mohali, Knowledge City, Sector 81, SAS Nagar, Manauli, Punjab 140306, India  }$}
\affiliation{\centerline{$^{a}$ Department of Physics, University of Mumbai, Mumbai, India}
\centerline{$^{b}$ UM-DAE Centre for Excellence in Basic Sciences (CEBS), Vidyanagari, Mumbai-400098, India }
 \centerline{$^{c}$ Adyar, Chennai }
\centerline{$^{d}$ Department of Theoretical Physics, Tata Institute of Fundamental Research,
Mumbai 400005, India }}
   \email{ $^\$ $ malvikadeo@hotmail.com, $^\dagger$ misra@physics.mu.ac.in, 
  $^{\$ \$}$ sharada.pbs@gmail.com,\\ $^{\dagger\dagger}$radhika.vinze@physics.mu.ac.in}
\begin{abstract}
In this article, we calculate the $S$-matrix elements for the process $e^{+} e^{-}\rightarrow t \bar{t}$ mediated by SM photon, $Z$ boson and an additional $Z^{'}$ boson indicating the contribution from new physics. We calculate the amplitude square using two component spinor formalism and four component spinor formalism and show the equivalance of the results using the two formalisms. We also establish the relations between the couplings of $Z^{'}$ boson to fermions in the two component spinor formalism and four component spinor formalism.
\end{abstract}
\maketitle
\section{Introduction}
There are many convincing reasons to look for the physics beyond Standard Model (SM) of particle physics eg. inability of the SM to explain the source of dark matter and dark energy or the anomalies between the predicted and experimental values of observables in flavour physics \cite{Crivellin_2024}. One of the proposals to go beyond the SM is to extend SM by an additional $U(1)$ symmetry. The additional symmetry results in a neutral $Z^{'}$ boson \cite{Langacker_2009} which gains its mass via Stuckelberg mechanism \cite{K_rs_2004, Vinze:2021ezd} or by addition of extra Higgs doublet which breaks the additional symmetry spontaneously giving mass to the $Z^{'}$. Since, $Z^{'}$ couples with the fermions in the SM, the resulting effects can account for the discrepancy in the theoretically predicted and observed values of certain parameters \cite{Rizzo:2006nw}. 

Since its discovery at CDF\cite{CDF:1995wbb} and D0\cite{D0:1995jca} experiments, the top quark has played an important role in the study of physics of the Standard Model (SM) as well as in the study of the physics beyond SM. It is the heaviest member of SM with largest Yukawa couplings. The coupling of the top quark with the Higgs boson is close to unity, as a result, the top quark plays a vital role in the higher order corrections to Higgs boson mass, thus, improving the understanding of electroweak sector of the SM. The uncertainty in the top quark properties such as top quark mass\cite{ATLAS:2018fwq, CMS:2015lbj}, top quark pair production cross section\cite{CMS:2018fks, ATLAS:2019hau}, its coupling with other quarks \cite{CMS:2020vac} can be interpreted as an evidence for physics beyond the SM\cite{Franceschini:2023nlp}. Hence, there is a need to measure top quark properties with better accuracy.

Electron positron colliders provide a clean environment for measuring top quark couplings with greater accuracy as compared to hadron colliders since at $e^{+} e^{-}$ collider there is no simultaneous QCD top quark production\cite{Amjad:2013tlv}. As top quark decays via electroweak interaction before hadronisation, there are significant angular correlations between the spin of the top quark and its decay products\cite{Parke:1996pr}. These angular correlations greatly impact the couplings of top quark hence, the information obtained from the angular correlations can be used to put more precise bounds on the uncertainties in the top quark properties at $e^{+} e^{-}$ collider. For such studies, the helicity basis is used to decompose the top quark spin\cite{Schmidt:1995mr}.

At very high energies, the particles in the initial and final state become effectively massless. As a result, the chirality and helicity of the particle at very high energies become the same. Thus, one can shift to the helicity basis for writing the amplitude for the scattering process at very high energies. The helicity amplitude method is based on the helicity basis and is useful when there are large number of particles in the final state. As the number of particles in the final state increases, the number of interference terms in the amplitude square increases exponentially and then it becomes difficult to compute amplitude square using four component Dirac spinors. Calculation of such processes becomes simpler when one uses helicity amplitude method. The wave functions of fermions in the helicity basis are two component Weyl spinors.  In two component spinor mechanism, the amplitude for the scattering process is written taking into account the left/right-handedness of the particle. Thus, the Feynman diagrams consider all the possible combinations of chiralities of the initial and final state particles. Similar to $4\times4~ \gamma$ matrices in four component spinor mechanism, the trace properties for $2\times2$ sigma matrices are used in two component spinor mechanism. The Feynman rules to compute amplitude square in two component formalism are summarised in \cite{Dreiner_2010}.

In this article, we present the detailed calculation of the $S$ matrix elements for the process $e^{+}$ $e^{-} \rightarrow t \Bar{t}$ at leading order in an extension of SM which has an additional massive $Z^{'}$ boson. In this model, the process under consideration can be mediated by three bosons, a massless photon ($\gamma$), a massive $Z$ boson or a massive $Z^{'}$ belonging to the new physics (NP) model. The calculation is performed using two component spinor formalism as well as the conventional four component spinor formalism. Though we include $Z^{'}$ contribution in this interaction, we implement a model independent approach for $Z^{'}$, hence the choice of couplings of $Z^{'}$ is not restricted. Our aim is to show the equivalence between two component spinor mechanism and four component spinor mechanism for spin averaged amplitude square of the process $e^{+}$ $e^{-} \rightarrow t \Bar{t}$ mediated by $\gamma, Z, Z^{'}$. To achieve this, we calculate each amplitude square term in detail.

In the process of establishing the equivalence, we infer how the vector and axial vector couplings of the $Z^{'}$ boson in four component spinor mechanism are related to those in two component spinor mechanism. Our calculation of the cross term $\mathcal{M}_{ZZ^{'}}$ confirms the relations between the couplings in the two formalisms.

The plan of the article is as follows: In Section 2, we compute the amplitude for the process $e^{+}$ $e^{-} \rightarrow t \Bar{t}$ \; using two component helicity spinors. In Section 3, we compute the same amplitude square terms using four component Dirac spinors. In Section 4, we obtain the relations between vector and axial couplings for $Z^{'}$ boson in the two formalisms using the results obtained in Section 2 and Section 3. To check consistency of these relations, we compute the cross term  $\mathcal{M}_{ZZ^{'}}$ in Section 4 and verify the relations between couplings obtained. Finally in Section 5, we summarize our results.
\vspace{-0.5cm}
\section{Amplitude using two component spinors}
\begin{center} 
\begin{figure}
\includegraphics[width=0.7\textwidth]{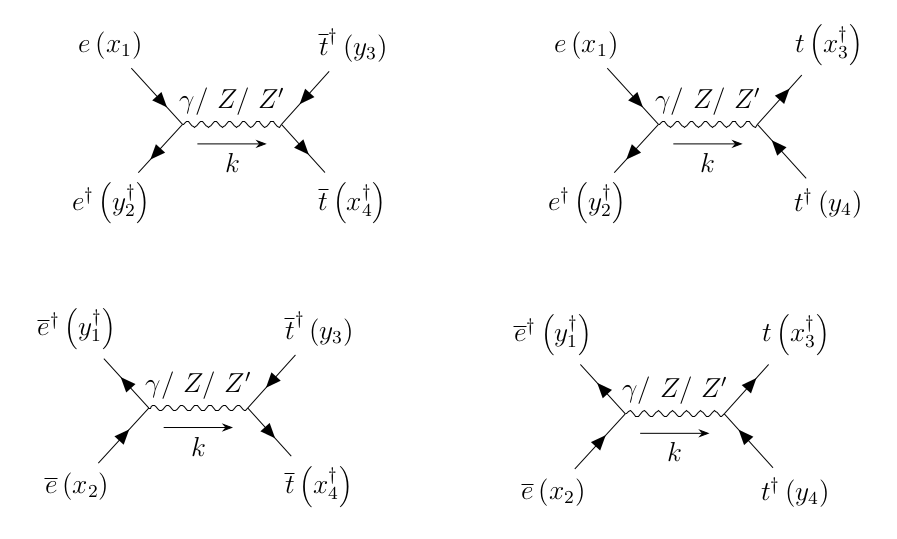} 
\caption{Feynman diagram for Top quark pair production in $e^{+}\;e^{-}$ scattering in helicity spinor formalism. The direction of the arrow defines the handedness of the particle. The wave-functions are assigned depending upon the handedness of the particle.}
\label{fig:twocomp}
\end{figure}
\end{center}

%%\label{}
%\vspace*{-0.2in} 
We consider scattering of $e^{-} e^{+}$ producing $t\Bar{t}$ pair mediated by SM photon ($\gamma$), $Z$ and NP $Z^{'}$. The wave-functions of the initial and final state particles depend upon the handedness of the particles, as a result, we get four terms in the amplitude corresponding to Feynman diagrams shown in Fig.\ref{fig:twocomp}. 
\vspace{-2.0cm}
Using the Feynman rules in Appendix \ref{FR-two}, the amplitude for $e^{-}\; e^{+} \rightarrow t \;\bar{t}$ mediated by photon ($\gamma$) exchange in two component spinor mechanism is
%\vspace{-0.2cm}
\begin{eqnarray}
	i\mathcal{M}_{\gamma} &=& \frac{-ig^{\mu\nu}}{k^2} \left[\left(iqQ_e\right)x_1\sigma_{\mu}y^{\dagger}_2 \left(-iqQ_t\right)y_{3}\sigma_{\nu} x^{\dagger}_{4} 	+ \left(iqQ_e\right) x_1\sigma_{\mu}y^{\dagger}_2 \left(-iqQ_t\right)x^{\dagger}_{3}\bar\sigma_{\nu}y_{4} \right.\nonumber\\
	&~& \left. +\left(iqQ_e\right)y^{\dagger}_{1}\bar\sigma_{\mu}x_{2}\left(-iqQ_t\right)y_{3}\sigma_{\nu} x^{\dagger}_{4} + \left(iqQ_e\right) y^{\dagger}_{1}\bar\sigma_{\mu}x_{2}\left(-iq Q_t \right) x^{\dagger}_{3}\bar\sigma_{\nu}y_{4}\right] \nonumber \\
 & = & \frac{-ig^{\mu\nu}}{k^2} \left[\left(iea_e\right)x_1\sigma_{\mu}y^{\dagger}_2 \left(-ie b_t\right)y_{3}\sigma_{\nu} x^{\dagger}_{4} 	+ \left(ie a_e\right) x_1\sigma_{\mu}y^{\dagger}_2 \left(-ie a_t\right)x^{\dagger}_{3}\bar\sigma_{\nu}y_{4} \right.\nonumber\\
	&~& \left. +\left(ie b_e\right)y^{\dagger}_{1}\bar\sigma_{\mu}x_{2}\left(-ie b_t\right)y_{3}\sigma_{\nu} x^{\dagger}_{4} + \left(ie b_e\right) y^{\dagger}_{1}\bar\sigma_{\mu}x_{2}\left(-iq a_t \right) x^{\dagger}_{3}\bar\sigma_{\nu}y_{4}\right] 
	\label{MG-two}
\end{eqnarray}
where $Q_{e}=-1$ is the charge of the electron and $Q_{t}= \frac{2}{3}$ is the charge of the top quark.
The amplitude for $e^{-}\; e^{+} \rightarrow t \;\bar{t}$ mediated by $Z$ boson is
\begin{eqnarray}
	i\mathcal{M}_{Z} &=&  \frac{i e^2}{ s_W^2 c_W^2 } \left(\frac{g_{\mu\nu} - \frac{k_{\mu} k_{\nu}}{M_{Z'}^2}} {k^2-M_{Z}^2}\right) \nonumber \\
 &~&\hspace{-0.7cm}\left[\;a_e b_t x_1\sigma_{\mu}y^{\dagger}_2 y_{3}\sigma_{\nu} x^{\dagger}_{4}  - a_e a_t x_1\sigma_{\mu}y^{\dagger}_2 x^{\dagger}_{3}\bar\sigma_{\nu}y_{4} 	 + b_e b_t y^{\dagger}_{1}\bar\sigma_{\mu}x_{2}y_{3}\sigma_{\nu} x^{\dagger}_{4}
	- b_e a_t y^{\dagger}_{1}\bar\sigma_{\mu}x_{2} x^{\dagger}_{3}\bar\sigma_{\nu}y_{4}\right]
	\label{MZ-two}
\end{eqnarray}
where $a_i, b_i~;~ i = e,t~$ are the respective electron and top quark vector and axial vector couplings with $Z$ given in Table \ref{Table1}.
\begin{table}
	\begin{tabular}{l c c c c} 
		\hline
		Mediator & $a_e$  & $b_e$ & $a_t$ & $b_t$\\ 
		\hline \vspace{-0.5cm} \\
		  $\gamma$  & $-1$ & $-1$ & $\frac{2}{3}$ & $\frac{2}{3}$ \vspace{0.2cm} \\ 
		$Z$ & $-\frac{1}{2} + s_{W}^{2}$ & $s_{W}^{2}$ & $\frac{1}{2}-\frac{2 s_{W}^{2}}{3}$	& $\frac{2 s_{W}^{2}}{3}$ \vspace{0.2cm} \\  
		$Z^{'}$ & $a_e^{'}$  & $b_e^{'}$ & $a_t^{'}$ & $b_t^{'}$  \\ 
		\hline
	\end{tabular}
	\caption{Couplings of $Z$, $Z^{'}$ with electron and top quark in two component spinor mechanism}
	\label{Table1}
\end{table} 
In a similar manner, the amplitude for $e^{-}\; e^{+} \rightarrow t \;\bar{t}$ mediated by $Z^{'}$ boson  is 
\begin{eqnarray}
	i\mathcal{M}_{Z'} &=& i \frac{ \; \eta_1 \eta_2 e^2}{ s_W^2 c_W^2 } \left(\frac{g_{\mu\nu} - \frac{k_{\mu} k_{\nu}}{M_{Z'}^2}} {k^2-M_{Z'}^2}\right) \nonumber \\
 &~&\hspace{-0.7cm}\left[a^{'}_e b^{'}_t x_1\sigma_{\mu}y^{\dagger}_2 y_{3}\sigma_{\nu} x^{\dagger}_{4}
	- a^{'}_e a^{'}_t x_1\sigma_{\mu}y^{\dagger}_2 x^{\dagger}_{3}\bar\sigma_{\nu}y_{4}  + b^{'}_e b^{'}_t y^{\dagger}_{1}\bar\sigma_{\mu}x_{2}y_{3}\sigma_{\nu} x^{\dagger}_{4} - b^{'}_e a^{'}_t y^{\dagger}_{1}\bar\sigma_{\mu}x_{2} x^{\dagger}_{3}\bar\sigma_{\nu}y_{4}\right]
	\label{MZp-two}
\end{eqnarray}
Here we introduce NP parameters $\eta_{1,2}$ indicating the difference in couplings of electron and top quark with $Z^{'}$ boson as compared to those with $Z$ boson. $a_i^{'}, b_i^{'}~;~ i = e,t~$ are the electron and top quark couplings with $Z^{'}$ given in Table \ref{Table1}. 
\\ 
Adding amplitudes corresponding to the three mediators, we get the total amplitude $\mathcal{M}$ for the process as, \vspace{-0.2cm}
\begin{equation}
	\mathcal{M}= \mathcal{M}_{\gamma} + \mathcal{M}_{Z}  + \mathcal{M}_{Z'}
 \vspace{-0.2cm}
\end{equation}
We rewrite $\left|\mathcal{M}\right|^2$ as,
\begin{eqnarray}
\vspace{-0.2cm}
	\left|\mathcal{M}\right|^2	&=& \left|\mathcal{M}_{\mathcal{D}}\right|^2 + \left|\mathcal{M}_{\mathcal{C}}\right|^2
\end{eqnarray}
where $\left|\mathcal{M}_{\mathcal{D}}\right|^2$ represents the sum of diagonal terms and $\left|\mathcal{M}_{\mathcal{C}}\right|^2$ represents sum of cross terms,
$$\left|\mathcal{M}_{\mathcal{D}}\right|^2 = \frac{1}{4} \sum_{i=\gamma, Z, Z^{'}}^{} \left|\mathcal{M}_{ii}\right|^2; \;\; \left|\mathcal{M}_{\mathcal{C}}\right|^2 = \sum_{\substack{i,j=\gamma, Z, Z^{'}\\i\neq j}}  \frac{1}{2}\sum \;\text{Re} \left[\mathcal{M}_{i}^\dagger \mathcal{M}_{j}\right]$$  
$\left|\mathcal{M}_{ii}\right|^2$ are calculated using eqns.\eqref{MG-two},\eqref{MZ-two} and \eqref{MZp-two}, and the two component spinor trace properties given in Appendix \ref{FR-two}. The results obtained are as follows -  
\begin{eqnarray}
	\frac{1}{4}\sum \mathcal{M}_{\gamma}^\dagger \mathcal{M}_{\gamma}  &=&  \frac{16 e^4 \left(2 p_1\cdot p_4 p_2\cdot p_3+2
   p_1\cdot p_3 p_2\cdot p_4+m_{t}^2 s\right)}{9 k^2}
	\label{MGMG-two}  \\ \nonumber \\
	\frac{1}{4}\sum \mathcal{M}_{Z}^\dagger \mathcal{M}_{Z}  &=& \frac{e^4}{2 c_{W}^4 s_{W}^4 \left(M_{Z}^3-k^2 M_{Z}\right)^2} \left[s \left(-2 s p_2\cdot p_3 \left(a_{e}^2 b_{t}^2+a_{t}^2 b_{e}^2\right)-2 s
   p_2\cdot p_4 \left(a_{e}^2 a_{t}^2 \right. \right. \right. \nonumber \\
   &~& \left. \left. \left.  \hspace{-2.5cm} +b_{e}^2 b_{t}^2\right)+\left(a_{e}^2+b_{e}^2\right)
   \left(a_{t}^2 s^2+4 a_{t} b_{t} m_{t}^2 M_{Z}^2+b_{t}^2 s^2\right)\right)-2 p_1\cdot p_4
   \left(s^2 \left(a_{e}^2 b_{t}^2+a_{t}^2 b_{e}^2\right)-2 p_2\cdot p_3 \right. \right. \nonumber \\
   &~& \left. \left.  \hspace{-2.5cm}\left(a_{e}^2
   \left(a_{t}^2 \left(2 M_{Z}^2-s\right)+b_{t}^2 s\right)+b_{e}^2 \left(a_{t}^2 s+b_{t}^2 \left(2 M_{Z}^2-s\right)\right)\right)\right) -2
   p_1\cdot p_3 \left(s^2 \left(a_{e}^2 a_{t}^2+b_{e}^2 b_{t}^2\right) \right. \right.  \nonumber \\
   &~& \left. \left.  \hspace{-2.5cm} -2
   p_2\cdot p_4 \left(a_{t}^2 \left(a_{e}^2 s+b_{e}^2 \left(2 M_{Z}^2-s\right)\right)+b_{t}^2
   \left(a_{e}^2 \left(2 M_{Z}^2-s\right)+b_{e}^2 s\right)\right)\right)\right] 
	\label{MZMZ-two-formula}
\end{eqnarray}
Substituting the couplings $a_e, b_e, a_t, b_t~$ corresponding to $Z$ boson from Table \ref{Table1}, we get	
\begin{eqnarray}
\frac{1}{4}\sum \mathcal{M}_{Z}^\dagger \mathcal{M}_{Z}  &=& \frac{e^4}{288 c_{W}^4 s_{W}^4 \left(M_{Z}^3-k^2
   M_{Z}\right)^2} \left(s \left(-8 s \left(32 s_{W}^4-40 s_{W}^2+13\right) s_{W}^4 p_2\cdot
   p_3 \right. \right. \nonumber \\
   &~& \left. \left. \hspace{-2.5cm} -2 s \left(128 s_{W}^8-160 s_{W}^6+148 s_{W}^4-60 s_{W}^2+9\right)
 p_2\cdot p_4 +\left(8 s_{W}^4-4 s_{W}^2+1\right) \left(16 m_{t}^2
   M_{Z}^2 s_{W}^2 \right. \right.  \right. \nonumber \\
   &~& \left. \left. \left. \hspace{-2.5cm} \left(4 s_{W}^2-3\right)+s^2 \left(32 s_{W}^4-24 s_{W}^2+9\right)\right)\right)-4
   p_1\cdot p_4 \left(\left(3 s \left(32 s_{W}^4-20 s_{W}^2+3\right)-2
   M_{Z}^2 \right. \right.  \right. \nonumber \\
   &~& \left. \left. \left. \hspace{-2.5cm} \left(128 s_{W}^8-160 s_{W}^6+148 s_{W}^4-60 s_{W}^2+9\right)\right) p_2\cdot
   p_3 +2 s^2 \left(32 s_{W}^4-40 s_{W}^2+13\right) s_{W}^4\right) \right. \nonumber \\
   &~&  \left. \hspace{-2.5cm} -2 p_1\cdot
   p_3 \left(s^2 \left(128 s_{W}^8-160 s_{W}^6+148 s_{W}^4-60 s_{W}^2+9\right)-2 \left(8
   M_{Z}^2 \left(32 s_{W}^4-40 s_{W}^2+13\right) s_{W}^4  \right. \right. \right. \nonumber \\
   &~&   \left. \left. \left. \hspace{-2.5cm} +s \left(96 s_{W}^4-60 s_{W}^2+9\right)\right)
  p_2\cdot p_4 \right)\right)
	\label{MZMZ-two}
\end{eqnarray}
The diagonal term for $Z^{'}$ boson results into 
\begin{eqnarray}
\frac{1}{4}\sum  \mathcal{M}_{Z'}^\dagger \mathcal{M}_{Z'} &=& \frac{e^4 \eta_{1}^2 \eta_{2}^2}{2 c_{W}^4
   s_{W}^4 \left(M_{Z'}^3-k^2 M_{Z'}\right)^2} \left[s \left(-2 s p_2\cdot p_3 \left({a_e^{'}}^2 {b_t^{'}}^2 + {a_t^{'}}^2
   {b_e^{'}}^2\right)-2 s p_2\cdot p_4 \left({a_e^{'}}^2 {a_t^{'}}^2 \right. \right. \right. \nonumber \\
   &~& \left. \left. \left.  \hspace{-2.5cm} + {b_e^{'}}^2
   {b_t^{'}}^2\right) +\left({a_e^{'}}^2+{b_e^{'}}^2\right) \left({a_t^{'}}^2 s^2+4 a_t^{'} b_t^{'} m_{t}^2 M_{Z'}^2+{b_t^{'}}^2
   s^2\right)\right)-2 p_1\cdot p_4 \left(s^2 \left({a_e^{'}}^2 {b_t^{'}}^2+{a_t^{'}}^2
   {b_e^{'}}^2\right) \right. \right. \nonumber \\
   &~& \left. \left.  \hspace{-2.5cm} -2 p_2\cdot p_3 \left({a_e^{'}}^2 \left({a_t^{'}}^2 \left(2
   M_{Z'}^2-s\right)+{b_t^{'}}^2 s\right)+{b_e^{'}}^2 \left({a_t^{'}}^2 s+{b_t^{'}}^2 \left(2 M_{Z'}^2-s\right)\right)\right)\right)-2
   p_1\cdot p_3 \left(s^2 \left({a_e^{'}}^2 {a_t^{'}}^2 \right. \right. \right. \nonumber \\
   &~& \left. \left. \left.  \hspace{-2.5cm} +{b_e^{'}}^2 {b_t^{'}}^2\right)-2
   p_2\cdot p_4 \left({a_t^{'}}^2 \left({a_e^{'}}^2 s+{b_e^{'}}^2 \left(2
   M_{Z'}^2-s\right)\right)+{b_t^{'}}^2 \left({a_e^{'}}^2 \left(2 M_{Z'}^2-s\right)+{b_e^{'}}^2 s\right)\right)\right)\right]
	\label{MZpMZp-two-formula}
\end{eqnarray}
The cross term between photon and $Z$ boson is
\begin{eqnarray}
\frac{1}{2}\sum \;\text{Re} \left|\mathcal{M}_{\gamma Z}\right|^2  &=&  -\frac{e^4}{3 c_{W}^2 k^2 M_{Z}^2 s_{W}^2
   \left(k^2-M_{Z}^2\right)} \nonumber \\
   &~& \hspace{-2.5cm} \left[-2 i s (a_{e}-b_{e}) (a_{t}+b_{t}) \bar{\epsilon }^{\overline{k}p_1p_2p_3}-2 i s
   (a_{e}-b_{e}) (a_{t}+b_{t}) \bar{\epsilon }^{\overline{k}p_1p_2p_4}+2 i a_{e} a_{t} s
   \bar{\epsilon }^{\overline{k}p_1p_3p_4}+2 i a_{e} a_{t} s \bar{\epsilon
   }^{\overline{k}p_2p_3p_4}  \right.  \nonumber \\
   &~& \left.  \hspace{-2.5cm} +16 a_{e} a_{t} M_{Z}^2 p_1\cdot
   p_4 p_2\cdot p_3-4 a_{e} a_{t} s p_1\cdot p_4 p_2\cdot p_3+4 a_{e} a_{t} s p_1\cdot
   p_3 p_2\cdot p_4-2 a_{e} a_{t} s^2 p_1\cdot p_3 \right.  \nonumber \\
   &~& \left.  \hspace{-2.5cm} -2 a_{e} a_{t} s^2 p_2\cdot p_4-2 i a_{e} b_{t} s \bar{\epsilon }^{\overline{k}p_1p_3p_4}-2 i a_{e} b_{t} s \bar{\epsilon
   }^{\overline{k}p_2p_3p_4}+16 a_{e} b_{t} M_{Z}^2 p_1\cdot p_3 p_2\cdot p_4+4 a_{e} b_{t} s p_1\cdot p_4 p_2\cdot p_3 \right.  \nonumber \\
   &~& \left.  \hspace{-2.5cm}  - 4 a_{e} b_{t} s p_1\cdot p_3 p_2\cdot p_4-2 a_{e} b_{t} s^2 p_1\cdot p_4-2 a_{e} b_{t} s^2 p_2\cdot p_3+2 i a_{t} b_{e} s \bar{\epsilon
   }^{\overline{k}p_1p_3p_4}+2 i a_{t} b_{e} s \bar{\epsilon
   }^{\overline{k}p_2p_3p_4} \right.  \nonumber \\
   &~& \left.  \hspace{-2.5cm} +16 a_{t} b_{e} M_{Z}^2 p_1\cdot p_3 p_2\cdot p_4+4 a_{t} b_{e} s p_1\cdot p_4 p_2\cdot p_3-4 a_{t} b_{e} s p_1\cdot p_3 p_2\cdot p_4 -2 a_{t} b_{e} s^2 p_1\cdot p_4  \right.  \nonumber \\
   &~& \left.  \hspace{-2.5cm}  -2 a_{t} b_{e} s^2 p_2\cdot p_3-2 i b_{e} b_{t} s \bar{\epsilon
   }^{\overline{k}p_1p_3p_4}-2 i b_{e} b_{t} s \bar{\epsilon
   }^{\overline{k}p_2p_3p_4}+16 b_{e} b_{t} M_{Z}^2 p_1\cdot
   p_4 p_2\cdot p_3-4 b_{e} b_{t} s p_1\cdot p_4 p_2\cdot p_3  \right.  \nonumber \\
   &~& \left.  \hspace{-2.5cm} +4 b_{e} b_{t} s p_1\cdot p_3 p_2\cdot p_4-2 b_{e} b_{t} s^2 p_1\cdot p_3 -2 b_{e} b_{t} s^2 p_2\cdot p_4+4 a_{e} a_{t} m_{t}^2 M_{Z}^2
   s \right.  \nonumber \\
   &~& \left.  \hspace{-2.5cm} +a_{e} a_{t} s^3+4 a_{e} b_{t} m_{t}^2 M_{Z}^2 s+a_{e} b_{t} s^3+4 a_{t} b_{e} m_{t}^2 M_{Z}^2 s+a_{t}
   b_{e} s^3+4 b_{e} b_{t} m_{t}^2 M_{Z}^2 s+b_{e} b_{t} s^3\right]
	\label{MGMZ-two-formula}
\end{eqnarray}
Substituting $a_e, b_e, a_t, b_t$ for the $Z$ boson from Table \ref{Table1}, we get
\begin{eqnarray}
\frac{1}{2}\sum \;\text{Re}  \left|\mathcal{M}_{\gamma Z}\right|^2 &=&  \frac{e^4}{36 c_{W}^2 k^2
   M_{Z}^2 s_{W}^2 \left(k^2-M_{Z}^2\right)} \nonumber \\
   &~& \hspace{-2.5cm} \left(2 i s \left(8 s_{W}^2-3\right) \bar{\epsilon }^{\overline{k}p_1p_2p_3}+2 i s \left(8
   s_{W}^2-3\right) \bar{\epsilon }^{\overline{k}p_1p_2p_4}-24 i s s_{W}^2 \bar{\epsilon
   }^{\overline{k}p_1p_3p_4}+6 i s \bar{\epsilon
   }^{\overline{k}p_1p_3p_4}-24 i s s_{W}^2 \bar{\epsilon
   }^{\overline{k}p_2p_3p_4} \right. \nonumber \\
   &~& \hspace{-2.5cm} \left. +6 i s \bar{\epsilon
   }^{\overline{k}p_2p_3p_4}+256 M_{Z}^2 s_{W}^4 p_1\cdot p_4 p_2\cdot p_3+256 M_{Z}^2 s_{W}^4 p_1\cdot p_3 p_2\cdot p_4  -160 M_{Z}^2 s_{W}^2 p_1\cdot p_4 p_2\cdot p_3 \right. \nonumber \\
   &~& \hspace{-2.5cm} \left. -160 M_{Z}^2 s_{W}^2 p_1\cdot p_3 p_2\cdot p_4+48 M_{Z}^2 p_1\cdot p_4
   p_2\cdot p_3-12 s p_1\cdot p_4 p_2\cdot p_3+12 s p_1\cdot p_3 p_2\cdot p_4 \right. \nonumber \\
   &~& \hspace{-2.5cm} \left. -32 s^2
   s_{W}^4 p_1\cdot p_3+20 s^2 s_{W}^2 p_1\cdot p_3-6 s^2
   p_1\cdot p_3-32 s^2 s_{W}^4 p_1\cdot p_4+20 s^2 s_{W}^2
   p_1\cdot p_4-32 s^2 s_{W}^4 p_2\cdot p_3  \right. \nonumber \\
   &~& \hspace{-2.5cm} \left.  +20 s^2 s_{W}^2
   p_2\cdot p_3-32 s^2 s_{W}^4 p_2\cdot p_4+20 s^2 s_{W}^2
   p_2\cdot p_4-6 s^2 p_2\cdot p_4+128 m_{t}^2 M_{Z}^2 s
   s_{W}^4-80 m_{t}^2 M_{Z}^2 s s_{W}^2  \right. \nonumber \\
   &~& \hspace{-2.5cm} \left.  +12 m_{t}^2 M_{Z}^2 s+32 s^3 s_{W}^4-20 s^3 s_{W}^2+3 s^3\right)
	\label{MGMZ-two}
\end{eqnarray}
The cross term between photon and $Z^{'}$ boson is 
\begin{eqnarray}
	\frac{1}{2}\sum \;\text{Re} \left[\mathcal{M}_{\gamma}^\dagger \mathcal{M}_{Z^{'}}\right]	&=& -\frac{e^4 \eta_{1} \eta_{2}}{3 c_{W}^2 k^2 M_{Z'}^2
   s_{W}^2 \left(k^2-M_{Z'}^2\right)} \nonumber \\
   &~& \hspace{-2.5cm}\left[-2 i s (a_e^{'}-b_e^{'}) (a_t^{'}+b_t^{'}) \bar{\epsilon }^{\overline{k}p_1p_2p_3}-2 i
   s (a_e^{'}-b_e^{'}) (a_t^{'}+b_t^{'}) \bar{\epsilon }^{\overline{k}p_1p_2p_4}+2 i a_e^{'}
   a_t^{'} s \bar{\epsilon }^{\overline{k}p_1p_3p_4} \right. \nonumber \\
   &~& \hspace{-2.5cm} \left. +2 i a_e^{'} a_t^{'} s \bar{\epsilon
   }^{\overline{k}p_2p_3p_4}+16 a_e^{'} a_t^{'} M_{Z'}^2 p_1\cdot p_4 p_2\cdot p_3-4 a_e^{'} a_t^{'} s p_1\cdot p_4 p_2\cdot p_3+4 a_e^{'} a_t^{'} s p_1\cdot p_3 p_2\cdot p_4\right. \nonumber \\
   &~& \left.\hspace{-2.5cm} -2 a_e^{'} a_t^{'} s^2 p_1\cdot p_3-2 a_e^{'} a_t^{'} s^2 p_2\cdot p_4-2 i a_e^{'} b_t^{'} s \bar{\epsilon
   }^{\overline{k}p_1p_3p_4}-2 i a_e^{'} b_t^{'} s \bar{\epsilon
   }^{\overline{k}p_2p_3p_4}+16 a_e^{'} b_t^{'} M_{Z'}^2 p_1\cdot p_3 p_2\cdot p_4\right. \nonumber \\
   &~& \left.\hspace{-2.5cm}+4 a_e^{'} b_t^{'} s p_1\cdot p_4 p_2\cdot p_3-4 a_e^{'} b_t^{'} s p_1\cdot p_3 p_2\cdot p_4-2 a_e^{'} b_t^{'} s^2 p_1\cdot p_4-2 a_e^{'} b_t^{'} s^2 p_2\cdot p_3+2 i a_t^{'} b_e^{'} s \bar{\epsilon
   }^{\overline{k}p_1p_3p_4}\right. \nonumber \\
   &~& \left.\hspace{-2.5cm} +2 i a_t^{'} b_e^{'} s \bar{\epsilon
   }^{\overline{k}p_2p_3p_4}+16 a_t^{'} b_e^{'} M_{Z'}^2 p_1\cdot p_3 p_2\cdot p_4+4 a_t^{'} b_e^{'} s p_1\cdot p_4 p_2\cdot p_3-4 a_t^{'} b_e^{'} s p_1\cdot p_3 p_2\cdot p_4\right. \nonumber \\
   &~& \left.\hspace{-2.5cm}-2 a_t^{'} b_e^{'} s^2 p_1\cdot p_4-2 a_t^{'} b_e^{'} s^2 p_2\cdot p_3-2 i b_e^{'} b_t^{'} s \bar{\epsilon
   }^{\overline{k}p_1p_3p_4}-2 i b_e^{'} b_t^{'} s \bar{\epsilon
   }^{\overline{k}p_2p_3p_4}+16 b_e^{'} b_t^{'} M_{Z'}^2 p_1\cdot p_4 p_2\cdot p_3\right. \nonumber 
   \end{eqnarray}
   \begin{eqnarray}
   &~& \left.-4 b_e^{'} b_t^{'} s p_1\cdot p_4 p_2\cdot p_3+4 b_e^{'} b_t^{'} s p_1\cdot p_3 p_2\cdot p_4-2 b_e^{'} b_t^{'} s^2 p_1\cdot p_3-2 b_e^{'} b_t^{'} s^2 p_2\cdot p_4+4 a_e^{'} a_t^{'} m_{t}^2
   M_{Z'}^2 s \right. \nonumber \\
   &~& \left. +a_e^{'} a_t^{'} s^3+4 a_e^{'} b_t^{'} m_{t}^2 M_{Z'}^2 s+a_e^{'} b_t^{'} s^3+4 a_t^{'} b_e^{'} m_{t}^2
   M_{Z'}^2 s+a_t^{'} b_e^{'} s^3+4 b_e^{'} b_t^{'} m_{t}^2 M_{Z'}^2 s+b_e^{'} b_t^{'} s^3\right] \nonumber \\
	\label{MGMZp-two-formula}
\end{eqnarray}	
The cross term between $Z$ boson and $Z^{'}$ boson is given in Appendix \ref{MZZp-two}.

In the next section, we calculate the same terms using four component spinor mechanism.

\section{Amplitude using four component spinors for SM mediators}
\label{}
In this section, we calculate the amplitude for $e^{+}e^{-} \rightarrow t \Bar{t}$ using four component Dirac spinors. Similar to calculations in the previous section, we have three terms corresponding to  mediators $\gamma, Z, Z^{'}$. We calculate diagonal terms and off-diagonal terms in $\left|\mathcal{M}\right|^2$. In this formalism, we calculate the spin averaged amplitude directly using Casimir trick and trace theorems, hence, we have only one diagram shown in Fig.\ref{fig:fourcomp} corresponding to the amplitude for each of the three vector boson exchanges. 
\begin{center} 
	\vspace{3.0cm}	
	\begin{figure}[h!]
		\centering
		% \vspace{-1.5cm}
		% trim=left botm right top.
		\includegraphics[trim=3.0cm 21cm 2.5cm 9cm, width=1.1\linewidth]{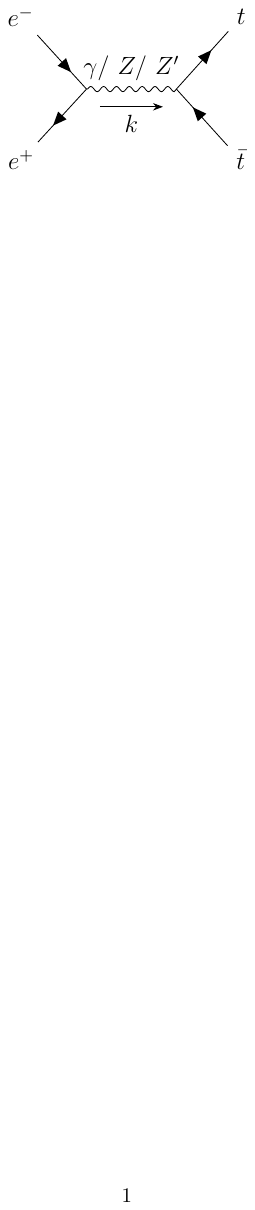} 
		\vspace*{-0.1in}  
		\caption{Feynman diagram for top quark pair production in $e^{+}\;e^{-}$ collision in four component spinor formalism}
		\label{fig:fourcomp}
  \vspace{-1cm}
	\end{figure}
\end{center} 
%\vspace{-1.5cm}
The amplitude for $e^{-}~e^{+} \rightarrow t~\bar{t}$ mediated by $\gamma$ in four component spinor mechanism with $Q_e = -1, Q_t = 2/3$ is
\begin{eqnarray}
i\mathcal{M}_{\gamma} &=& \overline v(p_2) \left(-iQ_{e}\gamma^{\mu}\right) u(p_1)  \left(\frac{-i~g_{\mu\nu}}{k^2}\right)  \overline u(p_3) \left(-iQ_{t}\gamma^{\nu}\right)  v(p_4) 
	\label{MG-four}
\end{eqnarray}
The amplitude for $e^{-}\; e^{+} \rightarrow t \;\bar{t}$ mediated by $Z$ boson in four component spinor mechanism is ($\sin{\theta_{W}} = s_W,  \cos{\theta_{W}} = c_W$)
\vspace{-0.2cm}
\begin{eqnarray}
\hspace{-0.4cm}i\mathcal{M}_{Z} &=& 
 \overline v(p_2) \left[\frac{-ie}{2s_W c_W}\gamma^{\mu}\left(C_{V}^e-C_{A}^e \gamma^5 \right)\right] u(p_1) \left(-i\frac{g_{\mu\nu} - \frac{k_{\mu} k_{\nu}}{M_{Z}^2}}{k^2-M_{Z}^2}\right) \overline u(p_3) \left[\frac{-ie}{2s_W c_W}\gamma^{\nu}\left(C_{V}^t-C_{A}^t \gamma^5 \right)\right] v(p_4) \nonumber \\
	\label{MZ-four}
\end{eqnarray}
where $C_{V}, C_{A}$ are the vector and axial vector couplings for $Z$ boson with electron and top quark as indicated by the superscript. These are listed in Table \ref{Table2}. 
\begin{table}
	\begin{tabular}{l c c c c} 
		\hline
		Mediator & $C_{V}^e$  & $C_{A}^e$ & $C_{V}^t$ & $C_{A}^t$\\ 
		\hline \vspace{-0.6cm} \\
		%  $\gamma$  & $1$ & $1$ & $\frac{2}{3}$ & $\frac{2}{3}$ \vspace{0.2cm} \\ 
		$Z$ & $2 s_{W}^2-\frac{1}{2}$ & $ -\frac{1}{2}$ & $\frac{1}{2}-\frac{4}{3} s_{W}^2$	& $\frac{1}{2}$ \vspace{0.1cm} \\  
		$Z^{'}$ & $C_{Ve}^{'}$  & $C_{A}^{e'}$ & $C_{V}^{t'}$ & $C_{A}^{t'}$  \\ 
		\hline
	\end{tabular}
	\caption{Couplings of $Z$, $Z^{'}$ with electron and top quark in four component spinor mechanism }
	\label{Table2}
 \vspace{-0.5cm}
\end{table} 
The amplitude for $e^{-}\; e^{+} \rightarrow t \;\bar{t}$ mediated by $Z^{'}$ boson in four component spinor mechanism is given as
%\vspace{0.5cm}
\begin{eqnarray}
\hspace{-0.5cm}i\mathcal{M}_{Z^{'}} &=&  \overline v(p_2) \left[\frac{-ie\eta_1}{2s_W c_W}\gamma^{\mu}\left(C_{V}^{'e}-C_{A}^{'e} \gamma^5 \right)\right] u(p_1)  \left(-i\frac{g_{\mu\nu} - \frac{k_{\mu} k_{\nu}}{M^2_{Z'}}}{k^2-M^2_{Z'}}\right) \overline u(p_3) \left[\frac{-ie\eta_2}{2s_W c_W}\gamma^{\nu}\left(C_{V}^{'t}-C_{A}^{'t} \gamma^5 \right)\right] v(p_4)  \nonumber \\
\label{MZp-four}
\end{eqnarray}
% \vspace{-2.5cm}
where $C_{V}^{'}, C_{A}^{'}$ are the vector and axial vector couplings for $Z^{'}$ boson with electron and top quark as indicated by the superscript.
\\
As we calculated the diagonal and off diagonal terms in two component spinor mechanism, in a similar manner we calculate the diagonal and off diagonal terms in four component spinor mechanism. Using trace properties of gamma matrices and eq. \eqref{MG-four}, \eqref{MZ-four} and \eqref{MZp-four}, we compute $\left|\mathcal{M}_{ii}\right|^2$ and $\left|\mathcal{M}_{ij}\right|^2$. We get the following results: 
\\
 %\vspace{-4.5cm}
\begin{eqnarray}
	\frac{1}{4}\sum \mathcal{M}_{\gamma}^\dagger \mathcal{M}_{\gamma} &=&\frac{16 e^4 \left(2 p_1\cdot p_4 p_2\cdot p_3+2
   p_1\cdot p_3 p_2\cdot p_4+m_{t}^2 s\right)}{9 k^2} \label{MGMG-four} \\  \nonumber \\
\frac{1}{4}\sum	\mathcal{M}_{Z}^\dagger \mathcal{M}_{Z}  \hspace{-0.1cm} &=&  \hspace{-0.1cm} -\frac{e^4}{8 c_{W}^4
   s_{W}^4 \left(M_{Z}^3-k^2 M_{Z}\right)^2} \left(s \left(s p_2\cdot p_3 \left(C_{Ae}^2
   \left(C_{At}^2+C_{Vt}^2\right) -4 C_{Ae} C_{At} C_{Ve} C_{Vt} \right. \right. \right.  \nonumber \\
   &~&  \left. \left. \left. \hspace{-2.5cm} +C_{Ve}^2
   \left(C_{At}^2+C_{Vt}^2\right)\right) +s p_2\cdot p_4 \left(C_{Ae}^2
   \left(C_{At}^2+C_{Vt}^2\right)+4 C_{Ae} C_{At} C_{Ve} C_{Vt}+C_{Ve}^2
   \left(C_{At}^2+C_{Vt}^2\right)\right) \right. \right.  \nonumber \\
   &~&  \left. \left. \hspace{-2.5cm}  +\left(C_{Ae}^2+C_{Ve}^2\right) \left(C_{At}^2 \left(2 m_{t}^2
   M_{Z}^2-s^2\right)-C_{Vt}^2 \left(2 m_{t}^2 M_{Z}^2+s^2\right)\right)\right)+\left(p_1\cdot
   p_4\right) \left(s^2 \left(C_{Ae}^2 \left(C_{At}^2+C_{Vt}^2\right)  \right. \right.  \right. \nonumber \\
   &~&  \left. \left. \left. \hspace{-2.5cm}  -4 C_{Ae} C_{At} C_{Ve}
   C_{Vt}+C_{Ve}^2 \left(C_{At}^2+C_{Vt}^2\right)\right)-4 p_2\cdot p_3
   \left(C_{Ae}^2 M_{Z}^2 \left(C_{At}^2+C_{Vt}^2\right)+4 C_{Ae} C_{At} C_{Ve} C_{Vt}
    \right. \right.  \right. \nonumber \\
   &~&  \left. \left. \left. \hspace{-2.5cm}  \left(M_{Z}^2-s\right)+C_{Ve}^2 M_{Z}^2
   \left(C_{At}^2+C_{Vt}^2\right)\right)\right)+p_1\cdot p_3 \left(s^2
   \left(C_{Ae}^2 \left(C_{At}^2+C_{Vt}^2\right)+4 C_{Ae} C_{At} C_{Ve} C_{Vt}   \right. \right.  \right. \nonumber \\
   &~&  \left. \left. \left. \hspace{-2.5cm}  +C_{Ve}^2
   \left(C_{At}^2+C_{Vt}^2\right)\right)-4 p_2\cdot p_4 \left(C_{Ae}^2
   M_{Z}^2 \left(C_{At}^2+C_{Vt}^2\right)+4 C_{Ae} C_{At} C_{Ve} C_{Vt}
   \left(s-M_{Z}^2\right)  \right. \right.  \right. \nonumber \\
   &~&  \left. \left. \left. \hspace{-2.5cm} +C_{Ve}^2 M_{Z}^2 \left(C_{At}^2+C_{Vt}^2\right)\right)\right)\right)
	\label{MZMZ-four-formula}
\end{eqnarray}
Substituting $C_V, C_A$ for $Z$ boson with electron and top quark from Table \ref{Table2}, and averaging over the spins of initial state and final states, we get
\begin{eqnarray}
	\hspace{-0.4cm}\frac{1}{4}\sum \mathcal{M}_{Z}^\dagger \mathcal{M}_{Z}  &=&   \frac{e^4}{288 c_{W}^4 s_{W}^4 \left(M_{Z}^3-k^2
   M_{Z}\right)^2} \nonumber \\
   &~& \hspace{-2.5cm} \left(s \left(-8 s \left(32 s_{W}^4-40 s_{W}^2+13\right) s_{W}^4 p_2\cdot p_3 -2 s \left(128 s_{W}^8-160 s_{W}^6+148 s_{W}^4-60 s_{W}^2+9\right)
   p_2\cdot p_4 \right. \right.  \nonumber \\
   &~& \hspace{-2.5cm} \left. \left. +\left(8 s_{W}^4-4 s_{W}^2+1\right) \left(16 m_{t}^2
   M_{Z}^2 s_{W}^2 \left(4 s_{W}^2-3\right)+s^2 \left(32 s_{W}^4-24 s_{W}^2+9\right)\right)\right)-4 p_1\cdot p_4 \right.  \nonumber \\
   &~& \hspace{-2.5cm} \left. \left(\left(3 s \left(32 s_{W}^4-20 s_{W}^2+3\right)-2
   M_{Z}^2 \left(128 s_{W}^8-160 s_{W}^6+148 s_{W}^4-60 s_{W}^2+9\right)\right) p_2\cdot p_3 \right. \right.  \nonumber \\
   &~& \hspace{-2.5cm} \left. \left. +2 s^2 \left(32 s_{W}^4-40 s_{W}^2+13\right) s_{W}^4\right)-2 p_1\cdot p_3 \left(s^2 \left(128 s_{W}^8-160 s_{W}^6+148 s_{W}^4-60 s_{W}^2+9\right) \right. \right.  \nonumber \\
   &~& \hspace{-2.5cm} \left. \left. -2 \left(8
   M_{Z}^2 \left(32 s_{W}^4-40 s_{W}^2+13\right) s_{W}^4+s \left(96 s_{W}^4-60 s_{W}^2+9\right)\right)
   p_2\cdot p_4\right)\right) \label{MZMZ-four} \\   \nonumber \\ 
\frac{1}{4}\sum \mathcal{M}_{Z'}^\dagger \mathcal{M}_{Z'} 	&=&	-\frac{e^4 \eta _{1}^2 \eta _{2}^2 }{8 c_{W}^4 s_{W}^4
   \left(M_{Z'}^3-k^2 M_{Z'}\right)^2} \nonumber \\
   &~& \hspace{-2.5cm}\left(s \left(s p_2\cdot p_3 \left(C_{Ae}^{'2}
   \left(C_{At}^{'2}+C_{Vt}^{'2}\right)-4 C_{Ae}^{'} C_{At}^{'} C_{Ve}^{'} C_{Vt}^{'}+C_{Ve}^{'2}
   \left(C_{At}^{'2}+C_{Vt}^{'2}\right)\right)  \right. \right.  \nonumber \\
   &~& \hspace{-2.5cm} \left. \left.  +s p_2\cdot p_4 \left(C_{Ae}^{'2}
   \left(C_{At}^{'2}+C_{Vt}^{'2}\right)+4 C_{Ae}^{'} C_{At}^{'} C_{Ve}^{'} C_{Vt}^{'}+C_{Ve}^{'2}
   \left(C_{At}^{'2}+C_{Vt}^{'2}\right)\right)+\left(C_{Ae}^{'2}+C_{Ve}^{'2}\right)  \right. \right.  \nonumber \\
   &~& \hspace{-2.5cm} \left. \left.  \left(C_{At}^{'2} \left(2 m_{t}^2
   M_{Z'}^2-s^2\right)-C_{Vt}^{'2} \left(2 m_{t}^2 M_{Z'}^2+s^2\right)\right)\right)+\left(p_1\cdot
   p_4\right) \left(s^2 \left(C_{Ae}^{'2} \left(C_{At}^{'2}+C_{Vt}^{'2}\right)  \right. \right. \right.  \nonumber \\
   &~& \hspace{-2.5cm} \left. \left. \left. -4 C_{Ae}^{'} C_{At}^{'} C_{Ve}^{'}
   C_{Vt}^{'}+C_{Ve}^{'2} \left(C_{At}^{'2}+C_{Vt}^{'2}\right)\right)-4 p_2\cdot p_3
   \left(C_{Ae}^{'2} M_{Z'}^2 \left(C_{At}^{'2}+C_{Vt}^{'2}\right)+4 C_{Ae}^{'} C_{At}^{'} C_{Ve}^{'} C_{Vt}^{'}
    \right. \right. \right.  \nonumber \\
   &~& \hspace{-2.5cm} \left. \left. \left.  \left(M_{Z'}^2-s\right)+C_{Ve}^{'2} M_{Z'}^2 \left(C_{At}^{'2}+C_{Vt}^{'2}\right)\right)\right)+\left(p_1\cdot
   p_3\right) \left(s^2 \left(C_{Ae}^{'2} \left(C_{At}^{'2}+C_{Vt}^{'2}\right)+4 C_{Ae}^{'} C_{At}^{'} C_{Ve}^{'}
   C_{Vt}^{'}  \right. \right. \right.  \nonumber \\
   &~& \hspace{-2.5cm} \left. \left. \left. +C_{Ve}^{'2} \left(C_{At}^{'2}+C_{Vt}^{'2}\right)\right)-4 p_2\cdot p_4
   \left(C_{Ae}^{'2} M_{Z'}^2 \left(C_{At}^{'2}+C_{Vt}^{'2}\right)+4 C_{Ae}^{'} C_{At}^{'} C_{Ve}^{'} C_{Vt}^{'}
   \left(s-M_{Z'}^2\right)   \right. \right. \right.  \nonumber \\
   &~& \hspace{-2.5cm} \left. \left. \left. +C_{Ve}^{'2} M_{Z'}^2 \left(C_{At}^{'2}+C_{Vt}^{'2}\right)\right)\right)\right)
	\label{MZpMZp-four-formula}
 \vspace{-0.5cm}
\end{eqnarray}
Now, we calculate the cross terms $\mathcal{M}_{\gamma}^\dagger \mathcal{M}_{Z}$ and $\mathcal{M}_{\gamma}^\dagger \mathcal{M}_{Z^{'}}$ using four component spinor mechanism. Using the trace properties of gamma matrices, we get the following results -
\begin{eqnarray}
	\frac{1}{2}\sum \mathcal{M}_{\gamma}^\dagger \mathcal{M}_{Z}  &=& -\frac{e^4}{3 c_{W}^2 k^2 M_{Z}^2 s_{W}^2 \left(k^2-M_{Z}^2\right)} \left[8 C_{Ae} C_{At} M_{Z}^2 p_1\cdot p_4 p_2\cdot  p_3 -8 C_{Ae} C_{At} M_{Z}^2 p_1\cdot p_3 p_2\cdot p_4 \right. \nonumber \\
 &~& \hspace{-2.5cm} \left. -4 C_{Ae} C_{At} s p_1\cdot p_4 p_2\cdot p_3 +4 C_{Ae} C_{At} s p_1\cdot p_3 p_2\cdot  p_4 -C_{Ae} C_{At} s^2 p_1\cdot p_3+C_{Ae} C_{At} s^2
   p_1\cdot p_4  \right. \nonumber \\
 &~& \hspace{-2.5cm} \left.  +C_{Ae} C_{At} s^2 p_2\cdot p_3-C_{Ae}
   C_{At} s^2 p_2\cdot p_4-2 i C_{Ae} C_{Vt} s \bar{\epsilon
   }^{\overline{k}p_1p_2p_3}-2 i C_{Ae} C_{Vt} s \bar{\epsilon
   }^{\overline{k}p_1p_2p_4}+2 i C_{At} C_{Ve} s \bar{\epsilon
   }^{\overline{k}p_1p_3p_4}  \right. \nonumber \\
 &~& \hspace{-2.5cm} \left.  +2 i C_{At} C_{Ve} s \bar{\epsilon
   }^{\overline{k}p_2p_3p_4}+8 C_{Ve} C_{Vt} M_{Z}^2 p_1\cdot p_4 p_2\cdot p_3+8 C_{Ve} C_{Vt} M_{Z}^2 p_1\cdot p_3 p_2\cdot p_4-C_{Ve} C_{Vt} s^2 p_1\cdot p_3  \right. \nonumber \\
 &~& \hspace{-2.5cm} \left. -C_{Ve} C_{Vt} s^2 p_1\cdot p_4-C_{Ve} C_{Vt} s^2
   p_2\cdot p_3-C_{Ve} C_{Vt} s^2 p_2\cdot p_4+4 C_{Ve}
   C_{Vt} m_{t}^2 M_{Z}^2 s+C_{Ve} C_{Vt} s^3\right]
	\label{MGMZ-four-formula}
\end{eqnarray}
Substituting  $C_V, C_A$ for $Z$ boson with electron and top quark from Table \ref{Table2}, we get
\begin{eqnarray}
\frac{1}{2}\sum \;\text{Re}  \left|\mathcal{M}_{\gamma Z}\right|^2 &=&  \frac{e^4 }{36 c_{W}^2 k^2
   M_{Z}^2 s_{W}^2 \left(k^2-M_{Z}^2\right)} \nonumber \\
   &~& \hspace{-2.5cm} \left[2 i s \left(8 s_{W}^2-3\right) \bar{\epsilon }^{\overline{k}p_1p_2p_3}+2 i s \left(8
   s_{W}^2-3\right) \bar{\epsilon }^{\overline{k}p_1p_2p_4}-24 i s s_{W}^2 \bar{\epsilon
   }^{\overline{k}p_1p_3p_4} \right. \nonumber \\
 &~& \hspace{-2.5cm} \left.  +6 i s \bar{\epsilon
   }^{\overline{k}p_1p_3p_4}-24 i s s_{W}^2 \bar{\epsilon
   }^{\overline{k}p_2p_3p_4}+6 i s \bar{\epsilon
   }^{\overline{k}p_2p_3p_4}+256 M_{Z}^2 s_{W}^4 p_1\cdot p_4 p_2\cdot p_3+256 M_{Z}^2 s_{W}^4 p_1\cdot
   p_3 p_2\cdot p_4 \right. \nonumber \\
 &~& \hspace{-2.5cm} \left. -160 M_{Z}^2 s_{W}^2 p_1\cdot p_4 p_2\cdot p_3-160 M_{Z}^2 s_{W}^2 p_1\cdot p_3 p_2\cdot p_4+48 M_{Z}^2 p_1\cdot p_4
   p_2\cdot p_3-12 s p_1\cdot p_4 p_2\cdot p_3  \right. \nonumber \\
 &~& \hspace{-2.5cm} \left. + 12 s p_1\cdot p_3 p_2\cdot p_4-32 s^2
   s_{W}^4 p_1\cdot p_3+20 s^2 s_{W}^2 p_1\cdot p_3-6 s^2
   p_1\cdot p_3-32 s^2 s_{W}^4 p_1\cdot p_4+20 s^2 s_{W}^2
   p_1\cdot p_4 \right. \nonumber \\
 &~& \hspace{-2.5cm} \left. -32 s^2 s_{W}^4 p_2\cdot p_3+20 s^2 s_{W}^2
   p_2\cdot p_3-32 s^2 s_{W}^4 p_2\cdot p_4+20 s^2 s_{W}^2
   p_2\cdot p_4-6 s^2 p_2\cdot p_4+128 m_{t}^2 M_{Z}^2 s
   s_{W}^4 \right. \nonumber \\
 &~& \hspace{-2.5cm} \left. -80 m_{t}^2 M_{Z}^2 s s_{W}^2+12 m_{t}^2 M_{Z}^2 s+32 s^3 s_{W}^4-20 s^3 s_{W}^2+3 s^3\right]
	\label{MGMZ-four}
\end{eqnarray} 
The cross term between photon and $Z^{'}$ boson is given as 
\begin{eqnarray}
	\frac{1}{2}\sum \;\text{Re} \left[\mathcal{M}_{\gamma}^\dagger \mathcal{M}_{Z^{'}}\right]	&=& -\frac{e^4 \eta_{1} \eta_{2}}{3 c_{W}^2 k^2 M_{Z'}^2 s_{W}^2 \left[k^2-M_{Z'}^2\right)} \nonumber \\
 &~& \hspace{-2.5cm} \left[8 C_{Ae}^{'} C_{At}^{'} M_{Z'}^2 p_1\cdot p_4 p_2\cdot
   p_3 -8 C_{Ae}^{'} C_{At}^{'} M_{Z'}^2 p_1\cdot p_3 p_2\cdot
   p_4 -4 C_{Ae}^{'} C_{At}^{'} s p_1\cdot p_4 p_2\cdot
   p_3 \right. \nonumber \\
 &~& \hspace{-2.5cm} \left. +4 C_{Ae}^{'} C_{At}^{'} s p_1\cdot p_3 p_2\cdot
   p_4 -C_{Ae}^{'} C_{At}^{'} s^2 p_1\cdot p_3+C_{Ae}^{'} C_{At}^{'} s^2
   p_1\cdot p_4+C_{Ae}^{'} C_{At}^{'} s^2 p_2\cdot p_3 \right. \nonumber \\
 &~& \hspace{-2.5cm} \left. -C_{Ae}^{'}
   C_{At}^{'} s^2 p_2\cdot p_4-2 i C_{Ae}^{'} C_{Vt}^{'} s \bar{\epsilon
   }^{\overline{k}p_1p_2p_3}-2 i C_{Ae}^{'} C_{Vt}^{'} s \bar{\epsilon
   }^{\overline{k}p_1p_2p_4}  +2 i C_{At}^{'} C_{Ve}^{'} s \bar{\epsilon
   }^{\overline{k}p_1p_3p_4} \right. \nonumber \\
 &~& \hspace{-2.5cm} \left. +2 i C_{At}^{'} C_{Ve}^{'} s \bar{\epsilon
   }^{\overline{k}p_2p_3p_4}+8 C_{Ve}^{'} C_{Vt}^{'} M_{Z'}^2 p_1\cdot
   p_4 p_2\cdot p_3+8 C_{Ve}^{'} C_{Vt}^{'} M_{Z'}^2 p_1\cdot
   p_3 p_2\cdot p_4\right. \nonumber \\
 &~& \hspace{-2.5cm} \left. -C_{Ve}^{'} C_{Vt}^{'} s^2 p_1\cdot
   p_3 -C_{Ve}^{'} C_{Vt}^{'} s^2 p_1\cdot p_4-C_{Ve}^{'} C_{Vt}^{'} s^2
   p_2\cdot p_3-C_{Ve}^{'} C_{Vt}^{'} s^2 p_2\cdot p_4\right. \nonumber \\
 &~& \hspace{-2.5cm} \left.+4 C_{Ve}^{'}
   C_{Vt}^{'} m_{t}^2 M_{Z'}^2 s+C_{Ve}^{'} C_{Vt}^{'} s^3\right]
	\label{MGMZp-four-formula}
\end{eqnarray}
The cross term between $Z$ and $Z^{'}$ boson is given in Appendix \ref{MZMZp-four}

We have calculated all the terms in $\Bar{\left|\mathcal{M}\right|}^{2}$ using four component spinor mechanism. We can see from eq.\eqref{MGMG-two} and eq.\eqref{MGMG-four} that the terms calculated on the right hand side match exactly. Similarly, when we compare eq.\eqref{MZMZ-two} and eq.\eqref{MZMZ-four}, the right-hand side of both equations is identical. Thus, diagonal terms $\mathcal{M}_{\gamma}^\dagger \mathcal{M}_{\gamma}$ and $\mathcal{M}_{Z}^\dagger \mathcal{M}_{Z}$ match exactly in four component spinor mechanism and two component spinor mechanism. 
Similarly, comparing the cross term  $\mathcal{M}_{\gamma}^\dagger \mathcal{M}_{Z}$ from eq.\eqref{MGMZ-two} and eq.\eqref{MGMZ-four}, we get exactly matching result.

After verifying the equivalence for SM contribution, in the next section we compare the results of amplitude square terms obtained for NP $Z^{'}$ boson in two component and four component spinor formalisms and by imposing the equivalence, we derive the relations between the couplings of $Z^{'}$ boson in the two formalisms.

\section{Comparison of the amplitude terms}
We have obtained the amplitude square terms in the form of the vector and axial vector couplings for $Z^{'}$ boson in the previous sections. Since, the amplitude squares in two component and four component formalisms are equivalent for $e^{+} e^{-} \rightarrow t \bar{t}$ for $\gamma$ and $Z$ boson, we compare the amplitude squares for NP $Z^{'}$ boson in term by term manner so as to get the relations between the couplings in two component spinor formalism and four component spinor formalism. For simplicity, we consider $\eta_1 = \eta_2 = 1$~.

Firstly, we compare the result of diagonal term for $Z^{'}$ boson given by eqs. \eqref{MZpMZp-two-formula} and \eqref{MZpMZp-four-formula}. We compare the coefficients corresponding to $s^{2} p_2\cdot p_3, s^{2} p_2\cdot p_4, s^{2} p_1\cdot p_4$ and $ s^{2} p_1\cdot p_3  $ in both the equations, and get the following relations between the couplings:
%\begin{eqnarray}
 % 8 {a_e^{'}}^{2} {b_t^{'}}^{2} + 8 {a_t^{'}}^{2} {b_e^{'}}^{2} & =&  \left(C_{Ae}^{'2} + C_{Ve}^{'2} \right)  \left(C_{At}^{'2} + C_{Vt}^{'2}\right) - 4 C_{Ae}^{'} C_{At}^{'} C_{Ve}^{'} C_{Vt}^{'}  \label{eq1} \\
%  8 {a_e^{'}}^{2} {a_t^{'}}^{2} + 8 {b_e^{'}}^{2} {b_t^{'}}^{2} &=& \left(C_{Ae}^{'2} + C_{Ve}^{'2} \right)  \left(C_{At}^{'2} + C_{Vt}^{'2}\right) + 4 C_{Ae}^{'} C_{At}^{'} C_{Ve}^{'} C_{Vt}^{'}  \label{eq2}
%\end{eqnarray}
%By comparing the coefficients of $M_{Z^{'}}^{2} p_1\cdot p_4  $ and $p_1\cdot p_3 p_1\cdot p_3 $, we obtain the same relations. We compare the coefficients corresponding to $ p_1\cdot p_4 p_2\cdot p_3 $ and $ p_1\cdot p_3 p_2\cdot p_4 $ in eqs. \eqref{MZpMZp-two-formula} and \eqref{MZpMZp-four-formula} the equations, thus, we get:
\begin{eqnarray}
  	8 {a_e^{'}}^2 {b_t^{'}}^2 + 8 {a_t^{'}}^2 {b_e^{'}}^2  = \left(C_{Ae}^{'2} + C_{Ve}^{'2} \right) &~&  \hspace{-0.7cm}\left(C_{At}^{'2} + C_{Vt}^{'2}\right)  - 4 C_{Ae}^{'} C_{At}^{'} C_{Ve}^{'} C_{Vt}^{'}\; \; \label{eq1} \\
	8 {a_e^{'}}^2 {a_t^{'}}^2 + 8 {b_e^{'}}^2 {b_t^{'}}^2 = \left(C_{Ae}^{'2} + C_{Ve}^{'2}\right) &~&  \hspace{-0.7cm}\left(C_{At}^{'2} + C_{Vt}^{'2}\right) + 4 C_{Ae}^{'} C_{At}^{'} C_{Ve}^{'} C_{Vt}^{'}\; \; \label{eq2}\\
  4 \left({a_e^{'}}^2 {a_t^{'}}^2+ {b_e^{'}}^2 {b_t^{'}}^2\right) \left(2 M_{Z}^2-s\right)+ 4 s\left({a_e^{'}}^2 {b_t^{'}}^2+  {a_t^{'}}^2 {b_e^{'}}^2\right)   &=& \nonumber \\  M_{Z'}^2 \left(C_{Ae}^{'2} +C_{Ve}^{'2} \right)   \left(C_{At}^{'2}+C_{Vt}^{'2}\right) &~& \hspace{-0.7cm}+4 C_{Ae}^{'} C_{At}^{'} C_{Ve}^{'} C_{Vt}^{'}
\left(M_{Z'}^2-s\right)
\label{eq3}
 \end{eqnarray}
Next we compare the the cross term $\mathcal{M}_{\gamma}^\dagger \mathcal{M}_{Z^{'}}$ from eq. \eqref{MGMZp-two-formula} and eq. \eqref{MGMZp-four-formula} and get the relations:
\begin{eqnarray}
C_{Ve}^{'} C_{Vt}^{'} + C_{Ae}^{'} C_{At}^{'} & = & 2 a_e^{'} a_t^{'} + 2 b_e^{'}   b_t^{'}	\nonumber \\
C_{Ve}^{'} C_{Vt}^{'} - C_{Ae}^{'} C_{At}^{'} &=& 2 a_t^{'}   b_e^{'} + 2 a_e^{'} b_t^{'} \nonumber \\
C_{Ve}^{'} C_{Vt}^{'} &=& b_e^{'} b_t^{'} + a_e^{'} a_t^{'} + a_t^{'}   b_e^{'} + a_e^{'} b_t^{'}  
%C_{Ae}^{'} C_{At}^{'} &=& 
\label{eq4}
\end{eqnarray}
The solution to above equations \eqref{eq1}, \eqref{eq2}, \eqref{eq3} and \eqref{eq4} is given by -
\begin{eqnarray}
   \hspace{1cm} C_{Ve}^{'} &=& a_{e}^{'} + b_{e}^{'} \hspace{2cm} C_{Ae}^{'} = a_{e}^{'} - b_{e}^{'} \nonumber \\
   \hspace{1cm} C_{Vt}^{'} &=& a_{t}^{'} + b_{t}^{'} \hspace{2cm} C_{At}^{'} = a_{t}^{'} - b_{t}^{'}
    \label{soln}
\end{eqnarray}
Thus, we have obtained the relations between the vector and axial vector couplings for $Z^{'}$ boson in four component spinor formalism and in two component spinor formalism. Our next step is to check the consistency of these relations using the cross-term $\mathcal{M}_{Z}^\dagger \mathcal{M}_{Z'}$.
%\vspace{-3.2cm}

Since, we have obtained the relations between the $Z^{'}$ couplings in the two component spinor mechanism and in four component spinor mechanism in eq.\eqref{soln}, using these relations we compare result obtained for the cross term $\mathcal{M}_{Z}^\dagger \mathcal{M}_{Z^{'}}$ in four component formalism (using eqs.\eqref{MZ-four} and \eqref{MZp-four}) with the result obtained in two component formalism (using eqs. \eqref{MZ-two} and \eqref{MZp-two}).

Substituting eq. \eqref{soln} in eq.\eqref{MZZp-four-formula}, we get the result in eq.\eqref{MZMZp-two-formula}. When we compute the amplitude square for the cross term $\mathcal{M}_{Z}^\dagger \mathcal{M}_{Z^{'}}$ in two component spinor formalism from eqs. \eqref{MZ-two} and \eqref{MZp-two}, and substitute values of $a$ and $b$ for electron and top quark from Table \ref{Table1}, we get exactly the same result in eq. \eqref{MZMZp-two-formula}. Thus the relations obtained for vector and axial vector couplings are verified.

\section{Conclusion}
In this article, we have shown the equivalence between the two component spinor and four component spinor expressions of $S$ matrix elements for the process $e^{-} e^{+} \rightarrow t \Bar{t}$, taking into account scattering mediated by three neutral bosons namely standard photon, Z and a NP $Z^{'}$. We start with arbitrary couplings for $Z^{'}$ boson assuming same $C_{V}$-$C_A$ structure as in SM. While calculating the $\Bar{\left|\mathcal{M}\right|^2}$, we do not restrict the couplings of $Z^{'}$ boson to a specific model. By requiring the equivalence between both formalisms, we work out the relations between vector and axial vector couplings of electron and top quark to $Z^{'}$ in both the mechanisms. We also do not consider any specific mass range of $Z^{'}$ boson, thus, we obtain relations between vector and axial vector couplings in a model independent manner. 
The NP $Z^{'}$ boson considered in the calculation of the process under consideration can contribute to many NP scenarios such as flavour changing neutral currents of top quark\cite{Fuyuto_2016}. In this calculation, we have assumed that the $Z^{'}$ boson couples to the fermions exactly in the same manner as $Z$ boson does. It will be interesting to investigate these calculations in case of a more general couplings and also for specific NP models having $Z^{'}$ boson.

\section{Acknowledgements}
AM would like to thank Department of Atomic Energy, India for the award of Raja Ramanna Fellowship.

\newpage
\begin{appendices}
\renewcommand{\theequation}{A.\arabic{equation}}
\setcounter{equation}{0}
\section{Feynman rules and trace relations for two component spinor mechanism}
 \label{FR-two}
\renewcommand{\theequation}{A.\arabic{equation}}
\setcounter{equation}{0}
\begin{itemize}
\item Propagators for internal lines:
\item[] Photon :  $\Large{\frac{-ig^{\mu\nu}}{k^2}}$
\hspace{2cm} $Z$ boson : $ -\frac{i}{ k^2-M_{Z}^2 } \left(g_{\mu\nu} - \frac{k_{\mu} k_{\nu}} {k^2-M_{Z}^2}\right)$
\item Vertex factor:
\vspace{-0.5cm}
\item[] Photon ($Q_f $ is the charge of the fermion $f$) -  
\item[]
\item[] 
\item[] \begin{figure}[h!] 
\vspace*{-3.1cm} 
\centerline{\includegraphics[width=0.9 \textwidth]{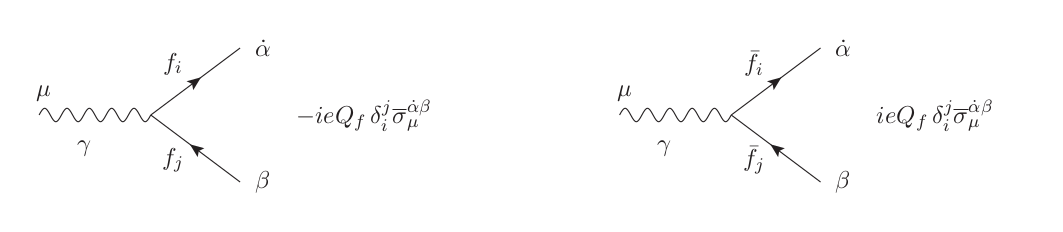} } \hspace*{-0.2in} 
\end{figure} 
\vspace{-3cm}
\item[] $Z$-boson - \\
($Q_f $ is the charge of the fermion $f$, $g s_{W}=e$, $T_3^f = \frac{1}{2}$ for top quark and $T_3^f = -\frac{1}{2}$ for electron. ) 
\item[] 
\item[]
\item[] \begin{figure}[h!] 
\vspace*{-3.1cm} 
\centerline{\includegraphics[width=0.9 \textwidth]{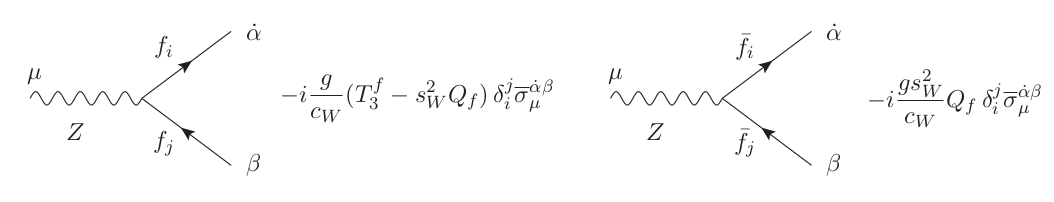} } \hspace*{-0.2in} 
\end{figure} 
\vspace{-3cm}
\item Identities - 
\vspace{-0.5cm}
\begin{eqnarray}
    x_{i} x_{i}^{\dagger} = y_{i} y_{i}^{\dagger} = p_{i}\cdot \sigma \hspace{3cm}
     x_{i}^{\dagger} x_{i} = y_{i}^{\dagger} y_{i}  = p_{i}\cdot \bar{\sigma} \nonumber \\
    x_{i} y_{i} = y_{i}^{\dagger} x_{i}^{\dagger}    = m_{i} \hspace{3cm}
  y_{i} x_{i}  = x_{i}^{\dagger} y_{i}^{\dagger}  = - m_{i} \nonumber   
\end{eqnarray}
\item The trace properties are -
\begin{eqnarray}
   \text{Tr}[\sigma^{\mu} \bar{\sigma}^\nu] &=& \text{Tr}[\bar{\sigma}^\mu\sigma^{\nu}] = 2 g^{\mu \nu}\nonumber \\
   \text{Tr}[\sigma^{\mu} \bar{\sigma}^\nu \sigma^{\rho} \bar{\sigma}^\kappa] &=&  = 2 [g^{\mu \nu} g^{\rho \kappa} - g^{\mu \rho} g^{\nu \kappa} + g^{\mu \kappa} g^{\nu \rho} + i\epsilon^{\mu \nu \rho \kappa}]\nonumber \\
    \text{Tr}[\bar{\sigma}^\mu \sigma^{\nu} \bar{\sigma}^\rho \sigma^{\kappa} ] &=&  = 2 [g^{\mu \nu} g^{\rho \kappa} - g^{\mu \rho} g^{\nu \kappa} + g^{\mu \kappa} g^{\nu \rho} - i\epsilon^{\mu \nu \rho \kappa}]\nonumber 
\end{eqnarray}
\end{itemize}
%%%%%%%%%%%%%%%%%%%%%%%%%%%%%%%%%%%%%%%%%%
\renewcommand{\theequation}{B.\arabic{equation}}
\setcounter{equation}{0}
\section{Cross term $M_{Z} M_{Z'}$ in two component spinor mechanism}
\label{MZZp-two}
 The cross term $\mathcal{M}_{Z}^\dagger \mathcal{M}_{Z^{'}}$ in two component spinor mechanism is
 \begin{eqnarray}
	\frac{1}{2}\sum \;\text{Re} \left[\mathcal{M}_{Z}^\dagger \mathcal{M}_{Z^{'}}\right]	&=& -\frac{e^4 \eta_{1} \eta_{2}}{48 c_{W}^4  s_{W}^4 M_{Z}^2 (k^2- M_{Z}^2) (k^2-M_{Z'}^2) M_{Z'}^2}  \nonumber \\
 &~& \hspace{-3.1cm} \left(8 a_{e}^{'}a_{t}^{'}M_{Z}^2 s^3 s_{W}^4 +8 a_{t}^{'}b_{e}^{'}M_{Z}^2 s^3 s_{W}^4+8 a_{e}^{'}b_{t}^{'}M_{Z}^2
   s^3 s_{W}^4+8 b_{e}^{'}b_{t}^{'}M_{Z}^2 s^3 s_{W}^4+8 a_{e}^{'}a_{t}^{'}M_{Z'}^2 s^3 s_{W}^4 \right. \nonumber \\
 &~& \left. \hspace{-3.1cm} +8 a_{t}^{'}b_{e}^{'}M_{Z'}^2 s^3
   s_{W}^4+8 a_{e}^{'}b_{t}^{'}M_{Z'}^2 s^3 s_{W}^4+8 b_{e}^{'}b_{t}^{'}M_{Z'}^2 s^3 s_{W}^4 +32 a_{e}^{'}a_{t}^{'}m_{t}^2
   M_{Z}^2 M_{Z'}^2 s s_{W}^4 \right. \nonumber \\
 &~& \left. \hspace{-3.1cm} +32 a_{t}^{'}b_{e}^{'}m_{t}^2 M_{Z}^2 M_{Z'}^2 s s_{W}^4+32 a_{e}^{'}b_{t}^{'}m_{t}^2 M_{Z}^2
   M_{Z'}^2 s s_{W}^4+32 b_{e}^{'}b_{t}^{'}m_{t}^2 M_{Z}^2 M_{Z'}^2 s s_{W}^4  \right. \nonumber \\
 &~& \left. \hspace{-3.1cm}  +16 i a_{e}^{'}a_{t}^{'}M_{Z}^2 s \bar{\epsilon
   }^{\overline{k}p_1p_2p_3} s_{W}^4-16 i a_{t}^{'}b_{e}^{'}M_{Z}^2 s \bar{\epsilon
   }^{\overline{k}p_1p_2p_3} s_{W}^4+16 i a_{e}^{'}b_{t}^{'}M_{Z}^2 s \bar{\epsilon
   }^{\overline{k}p_1p_2p_3} s_{W}^4  \right. \nonumber \\
 &~& \left. \hspace{-3.1cm}  -16 i b_{e}^{'}b_{t}^{'}M_{Z}^2 s \bar{\epsilon
   }^{\overline{k}p_1p_2p_3} s_{W}^4-16 i a_{e}^{'}a_{t}^{'}M_{Z'}^2 s \bar{\epsilon
   }^{\overline{k}p_1p_2p_3} s_{W}^4+16 i a_{t}^{'}b_{e}^{'}M_{Z'}^2 s \bar{\epsilon
   }^{\overline{k}p_1p_2p_3} s_{W}^4 \right. \nonumber \\
 &~& \left. \hspace{-3.1cm}  -16 i a_{e}^{'}b_{t}^{'}M_{Z'}^2 s \bar{\epsilon
   }^{\overline{k}p_1p_2p_3} s_{W}^4+16 i b_{e}^{'}b_{t}^{'}M_{Z'}^2 s \bar{\epsilon
   }^{\overline{k}p_1p_2p_3} s_{W}^4+16 i a_{e}^{'}a_{t}^{'}M_{Z}^2 s \bar{\epsilon
   }^{\overline{k}p_1p_2p_4} s_{W}^4  \right. \nonumber \\
 &~& \left. \hspace{-3.1cm} -16 i a_{t}^{'}b_{e}^{'}M_{Z}^2 s \bar{\epsilon
   }^{\overline{k}p_1p_2p_4} s_{W}^4+16 i a_{e}^{'}b_{t}^{'}M_{Z}^2 s \bar{\epsilon
   }^{\overline{k}p_1p_2p_4} s_{W}^4-16 i b_{e}^{'}b_{t}^{'}M_{Z}^2 s \bar{\epsilon
   }^{\overline{k}p_1p_2p_4} s_{W}^4 \right. \nonumber \\
 &~& \left. \hspace{-3.1cm} -16 i a_{e}^{'}a_{t}^{'}M_{Z'}^2 s \bar{\epsilon
   }^{\overline{k}p_1p_2p_4} s_{W}^4+16 i a_{t}^{'}b_{e}^{'}M_{Z'}^2 s \bar{\epsilon
   }^{\overline{k}p_1p_2p_4} s_{W}^4-16 i a_{e}^{'}b_{t}^{'}M_{Z'}^2 s \bar{\epsilon
   }^{\overline{k}p_1p_2p_4} s_{W}^4 \right. \nonumber \\
 &~& \left. \hspace{-3.1cm} +16 i b_{e}^{'}b_{t}^{'}M_{Z'}^2 s \bar{\epsilon
   }^{\overline{k}p_1p_2p_4} s_{W}^4-16 i a_{e}^{'}a_{t}^{'}M_{Z}^2 s \bar{\epsilon
   }^{\overline{k}p_1p_3p_4} s_{W}^4-16 i a_{t}^{'}b_{e}^{'}M_{Z}^2 s \bar{\epsilon
   }^{\overline{k}p_1p_3p_4} s_{W}^4 \right. \nonumber \\
 &~& \left. \hspace{-3.1cm} +16 i a_{e}^{'}b_{t}^{'}M_{Z}^2 s \bar{\epsilon
   }^{\overline{k}p_1p_3p_4} s_{W}^4+16 i b_{e}^{'}b_{t}^{'}M_{Z}^2 s \bar{\epsilon
   }^{\overline{k}p_1p_3p_4} s_{W}^4+16 i a_{e}^{'}a_{t}^{'}M_{Z'}^2 s \bar{\epsilon
   }^{\overline{k}p_1p_3p_4} s_{W}^4 \right. \nonumber \\
 &~& \left. \hspace{-3.1cm} +16 i a_{t}^{'}b_{e}^{'}M_{Z'}^2 s \bar{\epsilon
   }^{\overline{k}p_1p_3p_4} s_{W}^4-16 i a_{e}^{'}b_{t}^{'}M_{Z'}^2 s \bar{\epsilon
   }^{\overline{k}p_1p_3p_4} s_{W}^4-16 i b_{e}^{'}b_{t}^{'}M_{Z'}^2 s \bar{\epsilon
   }^{\overline{k}p_1p_3p_4} s_{W}^4 \right. \nonumber \\
 &~& \left. \hspace{-3.1cm}  -16 i a_{e}^{'}a_{t}^{'}M_{Z}^2 s \bar{\epsilon
   }^{\overline{k}p_2p_3p_4} s_{W}^4-16 i a_{t}^{'}b_{e}^{'}M_{Z}^2 s \bar{\epsilon
   }^{\overline{k}p_2p_3p_4} s_{W}^4+16 i a_{e}^{'}b_{t}^{'}M_{Z}^2 s \bar{\epsilon
   }^{\overline{k}p_2p_3p_4} s_{W}^4 \right. \nonumber \\
 &~& \left. \hspace{-3.1cm} +16 i b_{e}^{'}b_{t}^{'}M_{Z}^2 s \bar{\epsilon
   }^{\overline{k}p_2p_3p_4} s_{W}^4+16 i a_{e}^{'}a_{t}^{'}M_{Z'}^2 s \bar{\epsilon
   }^{\overline{k}p_2p_3p_4} s_{W}^4+16 i a_{t}^{'}b_{e}^{'}M_{Z'}^2 s \bar{\epsilon
   }^{\overline{k}p_2p_3p_4} s_{W}^4 \right. \nonumber \\
 &~& \left. \hspace{-3.1cm} -16 i a_{e}^{'}b_{t}^{'}M_{Z'}^2 s \bar{\epsilon
   }^{\overline{k}p_2p_3p_4} s_{W}^4-16 i b_{e}^{'}b_{t}^{'}M_{Z'}^2 s \bar{\epsilon
   }^{\overline{k}p_2p_3p_4} s_{W}^4-16 a_{e}^{'}a_{t}^{'}M_{Z}^2 s^2 p_1\cdot
   p_3 s_{W}^4  -16 b_{e}^{'}b_{t}^{'}M_{Z}^2 s^2 p_1\cdot p_3 s_{W}^4 \right. \nonumber \\
 &~& \left. \hspace{-3.1cm} -16
   a_{e}^{'}a_{t}^{'}M_{Z'}^2 s^2 p_1\cdot p_3 s_{W}^4-16 b_{e}^{'}b_{t}^{'}M_{Z'}^2 s^2
   p_1\cdot p_3 s_{W}^4  -16 a_{t}^{'}b_{e}^{'}M_{Z}^2 s^2 p_1\cdot
   p_4 s_{W}^4-16 a_{e}^{'}b_{t}^{'}M_{Z}^2 s^2 p_1\cdot p_4 s_{W}^4\right. \nonumber \\
 &~& \left. \hspace{-3.1cm}  -16
   a_{t}^{'}b_{e}^{'}M_{Z'}^2 s^2 p_1\cdot p_4 s_{W}^4  -16 a_{e}^{'}b_{t}^{'}M_{Z'}^2 s^2
   p_1\cdot p_4 s_{W}^4-16 a_{t}^{'}b_{e}^{'}M_{Z}^2 s^2 \left(p_2\cdot
   p_3\right) s_{W}^4-16 a_{e}^{'}b_{t}^{'}M_{Z}^2 s^2 p_2\cdot p_3 s_{W}^4\right. \nonumber \\
 &~& \left. \hspace{-3.1cm}-16
   a_{t}^{'}b_{e}^{'}M_{Z'}^2 s^2 p_2\cdot p_3 s_{W}^4 -16 a_{e}^{'}b_{t}^{'}M_{Z'}^2 s^2
   p_2\cdot p_3 s_{W}^4+128 a_{e}^{'}a_{t}^{'}M_{Z}^2 M_{Z'}^2 \left(p_1\cdot
   p_4\right) p_2\cdot p_3 s_{W}^4\right. \nonumber \\
 &~& \left. \hspace{-3.1cm}+128 b_{e}^{'}b_{t}^{'}M_{Z}^2 M_{Z'}^2
   p_1\cdot p_4 p_2\cdot p_3 s_{W}^4-32 a_{e}^{'}a_{t}^{'}
   M_{Z}^2 s p_1\cdot p_4 p_2\cdot p_3 s_{W}^4+32 a_{t}^{'}
   b_{e}^{'}M_{Z}^2 s p_1\cdot p_4 p_2\cdot p_3 s_{W}^4\right. \nonumber \\
 &~& \left. \hspace{-3.1cm}+32
   a_{e}^{'}b_{t}^{'}M_{Z}^2 s p_1\cdot p_4 p_2\cdot p_3
   s_{W}^4-32 b_{e}^{'}b_{t}^{'}M_{Z}^2 s p_1\cdot p_4 p_2\cdot
   p_3 s_{W}^4-32 a_{e}^{'}a_{t}^{'}M_{Z'}^2 s p_1\cdot p_4
   p_2\cdot p_3 s_{W}^4 \right. \nonumber \\
 &~& \left. \hspace{-3.1cm} +32 a_{t}^{'}b_{e}^{'}M_{Z'}^2 s p_1\cdot
   p_4 p_2\cdot p_3 s_{W}^4+32 a_{e}^{'}b_{t}^{'}M_{Z'}^2 s
   p_1\cdot p_4 p_2\cdot p_3 s_{W}^4-32 b_{e}^{'}b_{t}^{'}
   M_{Z'}^2 s p_1\cdot p_4 p_2\cdot p_3 s_{W}^4\right. \nonumber \\
 &~& \left. \hspace{-3.1cm}-16 a_{e}^{'}
   a_{t}^{'}M_{Z}^2 s^2 p_2\cdot p_4 s_{W}^4  -16 b_{e}^{'}b_{t}^{'}M_{Z}^2 s^2
   p_2\cdot p_4 s_{W}^4-16 a_{e}^{'}a_{t}^{'}M_{Z'}^2 s^2 p_2\cdot
   p_4 s_{W}^4\right. \nonumber \\
 &~& \left. \hspace{-3.1cm} -16 b_{e}^{'}b_{t}^{'}M_{Z'}^2 s^2 p_2\cdot p_4 s_{W}^4+128
   a_{t}^{'}b_{e}^{'}M_{Z}^2 M_{Z'}^2 p_1\cdot p_3 \left(p_2\cdot
   p_4\right) s_{W}^4+128 a_{e}^{'}b_{t}^{'}M_{Z}^2 M_{Z'}^2 p_1\cdot p_3
   p_2\cdot p_4 s_{W}^4 \right. \nonumber \\
   &~& \left. \hspace{-3.1cm}+32 a_{e}^{'}a_{t}^{'}M_{Z}^2 s p_1\cdot
   p_3 p_2\cdot p_4 s_{W}^4-32 a_{t}^{'}b_{e}^{'}M_{Z}^2 s
   p_1\cdot p_3 p_2\cdot p_4 s_{W}^4-32 a_{e}^{'}b_{t}^{'}
   M_{Z}^2 s p_1\cdot p_3 p_2\cdot p_4 s_{W}^4\right. \nonumber\\ \nonumber
   \end{eqnarray}
   \begin{eqnarray}
       &~& \left. +32 b_{e}^{'}
   b_{t}^{'}M_{Z}^2 s p_1\cdot p_3 p_2\cdot p_4 s_{W}^4+32
   a_{e}^{'}a_{t}^{'}M_{Z'}^2 s p_1\cdot p_3 p_2\cdot p_4
   s_{W}^4-32 a_{t}^{'}b_{e}^{'}M_{Z'}^2 s p_1\cdot p_3 \left(p_2\cdot
   p_4\right) s_{W}^4 \right. \nonumber \\
   &~& \left. -32 a_{e}^{'}b_{t}^{'}M_{Z'}^2 s p_1\cdot p_3
   p_2\cdot p_4 s_{W}^4+32 b_{e}^{'}b_{t}^{'}M_{Z'}^2 s \left(p_1\cdot
   p_3\right) p_2\cdot p_4 s_{W}^4-10 a_{e}^{'}a_{t}^{'}M_{Z}^2 s^3 s_{W}^2-6
   a_{t}^{'}b_{e}^{'}M_{Z}^2 s^3 s_{W}^2\right. \nonumber \\
   &~& \left.-4 a_{e}^{'}b_{t}^{'}M_{Z}^2 s^3 s_{W}^2-10 a_{e}^{'}a_{t}^{'}M_{Z'}^2 s^3 s_{W}^2-6
   a_{t}^{'}b_{e}^{'}M_{Z'}^2 s^3 s_{W}^2-4 a_{e}^{'}b_{t}^{'}M_{Z'}^2 s^3 s_{W}^2-16 a_{e}^{'}a_{t}^{'}m_{t}^2 M_{Z}^2
   M_{Z'}^2 s s_{W}^2\right. \nonumber \\
   &~& \left. -40 a_{e}^{'}b_{t}^{'}m_{t}^2 M_{Z}^2 M_{Z'}^2 s s_{W}^2-24 b_{e}^{'}b_{t}^{'}m_{t}^2 M_{Z}^2
   M_{Z'}^2 s s_{W}^2-20 i a_{e}^{'}a_{t}^{'}M_{Z}^2 s \bar{\epsilon }^{\overline{k}p_1p_2p_3}
   s_{W}^2+12 i a_{t}^{'}b_{e}^{'}M_{Z}^2 s \bar{\epsilon }^{\overline{k}p_1p_2p_3} s_{W}^2\right. \nonumber \\
   &~& \left.-8 i
   a_{e}^{'}b_{t}^{'}M_{Z}^2 s \bar{\epsilon }^{\overline{k}p_1p_2p_3} s_{W}^2+20 i a_{e}^{'}
   a_{t}^{'}M_{Z'}^2 s \bar{\epsilon }^{\overline{k}p_1p_2p_3} s_{W}^2-12 i a_{t}^{'}b_{e}^{'}
   M_{Z'}^2 s \bar{\epsilon }^{\overline{k}p_1p_2p_3} s_{W}^2+8 i a_{e}^{'}b_{t}^{'}M_{Z'}^2 s
   \bar{\epsilon }^{\overline{k}p_1p_2p_3} s_{W}^2\right. \nonumber \\
   &~& \left.-20 i a_{e}^{'}a_{t}^{'}M_{Z}^2 s \bar{\epsilon
   }^{\overline{k}p_1p_2p_4} s_{W}^2+12 i a_{t}^{'}b_{e}^{'}M_{Z}^2 s \bar{\epsilon
   }^{\overline{k}p_1p_2p_4} s_{W}^2-8 i a_{e}^{'}b_{t}^{'}M_{Z}^2 s \bar{\epsilon
   }^{\overline{k}p_1p_2p_4} s_{W}^2+20 i a_{e}^{'}a_{t}^{'}M_{Z'}^2 s \bar{\epsilon
   }^{\overline{k}p_1p_2p_4} s_{W}^2\right. \nonumber \\
   &~& \left.-12 i a_{t}^{'}b_{e}^{'}M_{Z'}^2 s \bar{\epsilon
   }^{\overline{k}p_1p_2p_4} s_{W}^2+8 i a_{e}^{'}b_{t}^{'}M_{Z'}^2 s \bar{\epsilon
   }^{\overline{k}p_1p_2p_4} s_{W}^2+20 i a_{e}^{'}a_{t}^{'}M_{Z}^2 s \bar{\epsilon
   }^{\overline{k}p_1p_3p_4} s_{W}^2+12 i a_{t}^{'}b_{e}^{'}M_{Z}^2 s \bar{\epsilon
   }^{\overline{k}p_1p_3p_4} s_{W}^2 \right. \nonumber \\
   &~& \left. -8 i a_{e}^{'}b_{t}^{'}M_{Z}^2 s \bar{\epsilon
   }^{\overline{k}p_1p_3p_4} s_{W}^2-20 i a_{e}^{'}a_{t}^{'}M_{Z'}^2 s \bar{\epsilon
   }^{\overline{k}p_1p_3p_4} s_{W}^2-12 i a_{t}^{'}b_{e}^{'}M_{Z'}^2 s \bar{\epsilon
   }^{\overline{k}p_1p_3p_4} s_{W}^2+8 i a_{e}^{'}b_{t}^{'}M_{Z'}^2 s \bar{\epsilon
   }^{\overline{k}p_1p_3p_4} s_{W}^2 \right. \nonumber \\
   &~& \left. +20 i a_{e}^{'}a_{t}^{'}M_{Z}^2 s \bar{\epsilon
   }^{\overline{k}p_2p_3p_4} s_{W}^2+12 i a_{t}^{'}b_{e}^{'}M_{Z}^2 s \bar{\epsilon
   }^{\overline{k}p_2p_3p_4} s_{W}^2-8 i a_{e}^{'}b_{t}^{'}M_{Z}^2 s \bar{\epsilon
   }^{\overline{k}p_2p_3p_4} s_{W}^2-20 i a_{e}^{'}a_{t}^{'}M_{Z'}^2 s \bar{\epsilon
   }^{\overline{k}p_2p_3p_4} s_{W}^2 \right. \nonumber \\
   &~& \left.  -12 i a_{t}^{'}b_{e}^{'}M_{Z'}^2 s \bar{\epsilon
   }^{\overline{k}p_2p_3p_4} s_{W}^2+8 i a_{e}^{'}b_{t}^{'}M_{Z'}^2 s \bar{\epsilon
   }^{\overline{k}p_2p_3p_4} s_{W}^2+20 a_{e}^{'}a_{t}^{'}M_{Z}^2 s^2 p_1\cdot
   p_3 s_{W}^2+20 a_{e}^{'}a_{t}^{'}M_{Z'}^2 s^2 p_1\cdot p_3 s_{W}^2 \right. \nonumber \\
   &~& \left. +12
   a_{t}^{'}b_{e}^{'}M_{Z}^2 s^2 p_1\cdot p_4 s_{W}^2+8 a_{e}^{'}b_{t}^{'}M_{Z}^2 s^2
   p_1\cdot p_4 s_{W}^2+12 a_{t}^{'}b_{e}^{'}M_{Z'}^2 s^2 p_1\cdot
   p_4 s_{W}^2+8 a_{e}^{'}b_{t}^{'}M_{Z'}^2 s^2 p_1\cdot p_4 s_{W}^2\right. \nonumber \\
   &~& \left.+12
   a_{t}^{'}b_{e}^{'}M_{Z}^2 s^2 p_2\cdot p_3 s_{W}^2+8 a_{e}^{'}b_{t}^{'}M_{Z}^2 s^2
   p_2\cdot p_3 s_{W}^2+12 a_{t}^{'}b_{e}^{'}M_{Z'}^2 s^2 p_2\cdot
   p_3 s_{W}^2+8 a_{e}^{'}b_{t}^{'}M_{Z'}^2 s^2 p_2\cdot p_3 s_{W}^2\right. \nonumber \\
   &~& \left.-160
   a_{e}^{'}a_{t}^{'}M_{Z}^2 M_{Z'}^2 p_1\cdot p_4 p_2\cdot
   p_3 s_{W}^2+40 a_{e}^{'}a_{t}^{'}M_{Z}^2 s p_1\cdot p_4
   p_2\cdot p_3 s_{W}^2-24 a_{t}^{'}b_{e}^{'}M_{Z}^2 s p_1\cdot
   p_4 p_2\cdot p_3 s_{W}^2\right. \nonumber \\
   &~& \left.-16 a_{e}^{'}b_{t}^{'}M_{Z}^2 s
   p_1\cdot p_4 p_2\cdot p_3 s_{W}^2+40 a_{e}^{'}a_{t}^{'}
   M_{Z'}^2 s p_1\cdot p_4 p_2\cdot p_3 s_{W}^2-24 a_{t}^{'}
   b_{e}^{'}M_{Z'}^2 s p_1\cdot p_4 p_2\cdot p_3 s_{W}^2 
 \right. \nonumber \\
   &~& \left.  -16
   a_{e}^{'}b_{t}^{'}M_{Z'}^2 s p_1\cdot p_4 p_2\cdot p_3
   s_{W}^2+20 a_{e}^{'}a_{t}^{'}M_{Z}^2 s^2 p_2\cdot p_4 s_{W}^2+20 a_{e}^{'}a_{t}^{'}M_{Z'}^2
   s^2 p_2\cdot p_4 s_{W}^2 \right. \nonumber \\
   &~& \left. -96 a_{t}^{'}b_{e}^{'}M_{Z}^2 M_{Z'}^2 p_1\cdot
   p_3 p_2\cdot p_4 s_{W}^2-64 a_{e}^{'}b_{t}^{'}M_{Z}^2 M_{Z'}^2
   p_1\cdot p_3 p_2\cdot p_4 s_{W}^2-40 a_{e}^{'}a_{t}^{'}
   M_{Z}^2 s p_1\cdot p_3 p_2\cdot p_4 s_{W}^2 \right. \nonumber \\
   &~& \left. +24 a_{t}^{'}
   b_{e}^{'}M_{Z}^2 s p_1\cdot p_3 p_2\cdot p_4 s_{W}^2+16
   a_{e}^{'}b_{t}^{'}M_{Z}^2 s p_1\cdot p_3 p_2\cdot p_4
   s_{W}^2-40 a_{e}^{'}a_{t}^{'}M_{Z'}^2 s p_1\cdot p_3 p_2\cdot
   p_4 s_{W}^2 \right. \nonumber \\
   &~& \left. +24 a_{t}^{'}b_{e}^{'}M_{Z'}^2 s p_1\cdot p_3
   p_2\cdot p_4 s_{W}^2+16 a_{e}^{'}b_{t}^{'}M_{Z'}^2 s p_1\cdot
   p_3 p_2\cdot p_4 s_{W}^2+3 a_{e}^{'}a_{t}^{'}M_{Z}^2 s^3+3 a_{e}^{'}a_{t}^{'}
   M_{Z'}^2 s^3 \right. \nonumber \\
   &~& \left. +12 a_{e}^{'}b_{t}^{'}m_{t}^2 M_{Z}^2 M_{Z'}^2 s+6 i a_{e}^{'}a_{t}^{'}M_{Z}^2 s \bar{\epsilon
   }^{\overline{k}p_1p_2p_3}-6 i a_{e}^{'}a_{t}^{'}M_{Z'}^2 s \bar{\epsilon
   }^{\overline{k}p_1p_2p_3}+6 i a_{e}^{'}a_{t}^{'}M_{Z}^2 s \bar{\epsilon
   }^{\overline{k}p_1p_2p_4} \right. \nonumber \\
   &~& \left. -6 i a_{e}^{'}a_{t}^{'}M_{Z'}^2 s \bar{\epsilon
   }^{\overline{k}p_1p_2p_4}-6 i a_{e}^{'}a_{t}^{'}M_{Z}^2 s \bar{\epsilon
   }^{\overline{k}p_1p_3p_4}+6 i a_{e}^{'}a_{t}^{'}M_{Z'}^2 s \bar{\epsilon
   }^{\overline{k}p_1p_3p_4}-6 i a_{e}^{'}a_{t}^{'}M_{Z}^2 s \bar{\epsilon
   }^{\overline{k}p_2p_3p_4} \right. \nonumber \\
   &~& \left. +6 i a_{e}^{'}a_{t}^{'}M_{Z'}^2 s \bar{\epsilon
   }^{\overline{k}p_2p_3p_4}-6 a_{e}^{'}a_{t}^{'}M_{Z}^2 s^2 \left(p_1\cdot
   p_3\right)-6 a_{e}^{'}a_{t}^{'}M_{Z'}^2 s^2 p_1\cdot p_3+48 a_{e}^{'}a_{t}^{'}
   M_{Z}^2 M_{Z'}^2 p_1\cdot p_4 p_2\cdot p_3 \right. \nonumber \\
   &~& \left. -12 a_{e}^{'}
   a_{t}^{'}M_{Z}^2 s p_1\cdot p_4 p_2\cdot p_3-12 a_{e}^{'}
   a_{t}^{'}M_{Z'}^2 s p_1\cdot p_4 p_2\cdot p_3-6 a_{e}^{'}
   a_{t}^{'}M_{Z}^2 s^2 p_2\cdot p_4-6 a_{e}^{'}a_{t}^{'}M_{Z'}^2 s^2 p_2\cdot
   p_4 \right. \nonumber \\
   &~& \left. +12 a_{e}^{'}a_{t}^{'}M_{Z}^2 s p_1\cdot p_3 p_2\cdot
   p_4+12 a_{e}^{'}a_{t}^{'}M_{Z'}^2 s p_1\cdot p_3 p_2\cdot
   p_4\right)
	\label{MZMZp-two-formula}
\end{eqnarray}
\renewcommand{\theequation}{C.\arabic{equation}}
\setcounter{equation}{0}
\section{Cross term $M_{Z} M_{Z'}$ in four component spinor mechanism}
The cross term $\mathcal{M}_{Z}^\dagger \mathcal{M}_{Z^{'}}$ in four component spinor mechanism is
\label{MZMZp-four}
 \begin{eqnarray}
	\frac{1}{2}\sum \;\text{Re} \left[\mathcal{M}_{Z}^\dagger \mathcal{M}_{Z^{'}}\right]	&=& -\frac{e^4 \eta_{1}\eta_{2} }{192 c_{W}^4 s_{W}^4M_{Z}^2 (k^2- M_{Z}^2) (k^2-M_{Z'}^2) M_{Z'}^2
    } \nonumber \\
    &~& \hspace{-3.1cm}\left(32 C_{Ve}^{'} C_{Vt}^{'} M_{Z}^2 s^3 s_{W}^4+32 C_{Ve}^{'} C_{Vt}^{'} M_{Z'}^2 s^3 s_{W}^4+128 C_{Ve}^{'} C_{Vt}^{'}
   m_t^2 M_{Z}^2 M_{Z'}^2 s s_{W}^4  \right. \nonumber \\
    &~& \hspace{-3.1cm} \left.  +64 i C_{Ae}^{'} C_{Vt}^{'} M_{Z}^2 s \bar{\epsilon
   }^{\overline{k}p_1p_2p_3} s_{W}^4-64 i C_{Ae}^{'} C_{Vt}^{'} M_{Z'}^2 s \bar{\epsilon
   }^{\overline{k}p_1p_2p_3} s_{W}^4+64 i C_{Ae}^{'} C_{Vt}^{'} M_{Z}^2 s \bar{\epsilon
   }^{\overline{k}p_1p_2p_4} s_{W}^4  \right. \nonumber \\
    &~& \hspace{-3.1cm} \left. -64 i C_{Ae}^{'} C_{Vt}^{'} M_{Z'}^2 s \bar{\epsilon
   }^{\overline{k}p_1p_2p_4} s_{W}^4-64 i C_{At}^{'} C_{Ve}^{'} M_{Z}^2 s \bar{\epsilon
   }^{\overline{k}p_1p_3p_4} s_{W}^4+64 i C_{At}^{'} C_{Ve}^{'} M_{Z'}^2 s \bar{\epsilon
   }^{\overline{k}p_1p_3p_4} s_{W}^4 \right. \nonumber \\
    &~& \hspace{-3.1cm} \left. -64 i C_{At}^{'} C_{Ve}^{'} M_{Z}^2 s \bar{\epsilon
   }^{\overline{k}p_2p_3p_4} s_{W}^4+64 i C_{At}^{'} C_{Ve}^{'} M_{Z'}^2 s \bar{\epsilon
   }^{\overline{k}p_2p_3p_4} s_{W}^4-32 C_{Ae}^{'} C_{At}^{'} M_{Z}^2 s^2 p_1\cdot
   p_3 s_{W}^4 \right. \nonumber \\
    &~& \hspace{-3.1cm} \left.  -32 C_{Ve}^{'} C_{Vt}^{'} M_{Z}^2 s^2 p_1\cdot p_3 s_{W}^4-32
   C_{Ae}^{'} C_{At}^{'} M_{Z'}^2 s^2 p_1\cdot p_3 s_{W}^4-32 C_{Ve}^{'} C_{Vt}^{'} M_{Z'}^2 s^2
   p_1\cdot p_3 s_{W}^4  \right. \nonumber \\     
   &~& \hspace{-3.1cm} \left. +32 C_{Ae}^{'} C_{At}^{'} M_{Z}^2 s^2 p_1\cdot
   p_4 s_{W}^4-32 C_{Ve}^{'} C_{Vt}^{'} M_{Z}^2 s^2 p_1\cdot p_4 s_{W}^4+32
   C_{Ae}^{'} C_{At}^{'} M_{Z'}^2 s^2 p_1\cdot p_4 s_{W}^4 \right. \nonumber \\     
   &~& \hspace{-3.1cm} \left. -32 C_{Ve}^{'} C_{Vt}^{'} M_{Z'}^2 s^2
   p_1\cdot p_4 s_{W}^4+32 C_{Ae}^{'} C_{At}^{'} M_{Z}^2 s^2 \left(p_2\cdot
   p_3\right) s_{W}^4-32 C_{Ve}^{'} C_{Vt}^{'} M_{Z}^2 s^2 p_2\cdot p_3 s_{W}^4 \right. \nonumber \\     
   &~& \hspace{-3.1cm} \left. +32
   C_{Ae}^{'} C_{At}^{'} M_{Z'}^2 s^2 p_2\cdot p_3 s_{W}^4-32 C_{Ve}^{'} C_{Vt}^{'} M_{Z'}^2 s^2
   p_2\cdot p_3 s_{W}^4 +256 C_{Ae}^{'} C_{At}^{'} M_{Z}^2 M_{Z'}^2 p_1\cdot p_4 p_2\cdot p_3 s_{W}^4  \right. \nonumber \\     
   &~& \hspace{-3.1cm} \left. +256 C_{Ve}^{'} C_{Vt}^{'} M_{Z}^2 M_{Z'}^2
   p_1\cdot p_4 p_2\cdot p_3 s_{W}^4-128 C_{Ae}^{'} C_{At}^{'}
   M_{Z}^2 s p_1\cdot p_4 p_2\cdot p_3 s_{W}^4 \right. \nonumber \\     
   &~& \hspace{-3.1cm} \left. -128 C_{Ae}^{'}
   C_{At}^{'} M_{Z'}^2 s p_1\cdot p_4 p_2\cdot p_3 s_{W}^4-32
   C_{Ae}^{'} C_{At}^{'} M_{Z}^2 s^2 p_2\cdot p_4 s_{W}^4-32 C_{Ve}^{'} C_{Vt}^{'} M_{Z}^2 s^2
   p_2\cdot p_4 s_{W}^4 \right. \nonumber \\     
   &~& \hspace{-3.1cm} \left. -32 C_{Ae}^{'} C_{At}^{'} M_{Z'}^2 s^2 \left(p_2\cdot
   p_4\right) s_{W}^4-32 C_{Ve}^{'} C_{Vt}^{'} M_{Z'}^2 s^2 p_2\cdot p_4 s_{W}^4-256
   C_{Ae}^{'} C_{At}^{'} M_{Z}^2 M_{Z'}^2 p_1\cdot p_3 p_2\cdot p_4 s_{W}^4 \right. \nonumber \\     
   &~& \hspace{-3.1cm} \left. +256 C_{Ve}^{'} C_{Vt}^{'} M_{Z}^2 M_{Z'}^2 p_1\cdot p_3
   p_2\cdot p_4 s_{W}^4 +128 C_{Ae}^{'} C_{At}^{'} M_{Z}^2 s p_1\cdot
   p_3 p_2\cdot p_4 s_{W}^4 \right. \nonumber \\     
   &~& \hspace{-3.1cm} \left. +128 C_{Ae}^{'} C_{At}^{'} M_{Z'}^2 s
   p_1\cdot p_3 p_2\cdot p_4 s_{W}^4-12 C_{At}^{'} C_{Ve}^{'}
   M_{Z}^2 s^3 s_{W}^2-8 C_{Ae}^{'} C_{Vt}^{'} M_{Z}^2 s^3 s_{W}^2 \right. \nonumber \\     
   &~& \hspace{-3.1cm} \left. -20 C_{Ve}^{'} C_{Vt}^{'} M_{Z}^2 s^3 s_{W}^2-12 C_{At}^{'} C_{Ve}^{'}
   M_{Z'}^2 s^3 s_{W}^2-8 C_{Ae}^{'} C_{Vt}^{'} M_{Z'}^2 s^3 s_{W}^2-20 C_{Ve}^{'} C_{Vt}^{'} M_{Z'}^2 s^3 s_{W}^2 \right. \nonumber \\     
   &~& \hspace{-3.1cm} \left. +48 C_{At}^{'}
   C_{Ve}^{'} m_t^2 M_{Z}^2 M_{Z'}^2 s s_{W}^2-32 C_{Ae}^{'} C_{Vt}^{'} m_t^2 M_{Z}^2 M_{Z'}^2 s s_{W}^2-80 C_{Ve}^{'}
   C_{Vt}^{'} m_t^2 M_{Z}^2 M_{Z'}^2 s s_{W}^2 \right. \nonumber \\     
   &~& \hspace{-3.1cm} \left. -24 i C_{Ae}^{'} C_{At}^{'} M_{Z}^2 s \bar{\epsilon
   }^{\overline{k}p_1p_2p_3} s_{W}^2-40 i C_{Ae}^{'} C_{Vt}^{'} M_{Z}^2 s \bar{\epsilon
   }^{\overline{k}p_1p_2p_3} s_{W}^2-16 i C_{Ve}^{'} C_{Vt}^{'} M_{Z}^2 s \bar{\epsilon
   }^{\overline{k}p_1p_2p_3} s_{W}^2 \right. \nonumber \\     
   &~& \hspace{-3.1cm} \left. +24 i C_{Ae}^{'} C_{At}^{'} M_{Z'}^2 s \bar{\epsilon
   }^{\overline{k}p_1p_2p_3} s_{W}^2+40 i C_{Ae}^{'} C_{Vt}^{'} M_{Z'}^2 s \bar{\epsilon
   }^{\overline{k}p_1p_2p_3} s_{W}^2+16 i C_{Ve}^{'} C_{Vt}^{'} M_{Z'}^2 s \bar{\epsilon
   }^{\overline{k}p_1p_2p_3} s_{W}^2 \right. \nonumber \\     
   &~& \hspace{-3.1cm} \left. -24 i C_{Ae}^{'} C_{At}^{'} M_{Z}^2 s \bar{\epsilon
   }^{\overline{k}p_1p_2p_4} s_{W}^2-40 i C_{Ae}^{'} C_{Vt}^{'} M_{Z}^2 s \bar{\epsilon
   }^{\overline{k}p_1p_2p_4} s_{W}^2-16 i C_{Ve}^{'} C_{Vt}^{'} M_{Z}^2 s \bar{\epsilon
   }^{\overline{k}p_1p_2p_4} s_{W}^2 \right. \nonumber \\     
   &~& \hspace{-3.1cm} \left. +24 i C_{Ae}^{'} C_{At}^{'} M_{Z'}^2 s \bar{\epsilon
   }^{\overline{k}p_1p_2p_4} s_{W}^2+40 i C_{Ae}^{'} C_{Vt}^{'} M_{Z'}^2 s \bar{\epsilon
   }^{\overline{k}p_1p_2p_4} s_{W}^2+16 i C_{Ve}^{'} C_{Vt}^{'} M_{Z'}^2 s \bar{\epsilon
   }^{\overline{k}p_1p_2p_4} s_{W}^2 \right. \nonumber \\     
   &~& \hspace{-3.1cm} \left. +16 i C_{Ae}^{'} C_{At}^{'} M_{Z}^2 s \bar{\epsilon
   }^{\overline{k}p_1p_3p_4} s_{W}^2+40 i C_{At}^{'} C_{Ve}^{'} M_{Z}^2 s \bar{\epsilon
   }^{\overline{k}p_1p_3p_4} s_{W}^2+24 i C_{Ve}^{'} C_{Vt}^{'} M_{Z}^2 s \bar{\epsilon
   }^{\overline{k}p_1p_3p_4} s_{W}^2 \right. \nonumber \\     
   &~& \hspace{-3.1cm} \left. -16 i C_{Ae}^{'} C_{At}^{'} M_{Z'}^2 s \bar{\epsilon
   }^{\overline{k}p_1p_3p_4} s_{W}^2-40 i C_{At}^{'} C_{Ve}^{'} M_{Z'}^2 s \bar{\epsilon
   }^{\overline{k}p_1p_3p_4} s_{W}^2-24 i C_{Ve}^{'} C_{Vt}^{'} M_{Z'}^2 s \bar{\epsilon
   }^{\overline{k}p_1p_3p_4} s_{W}^2 \right. \nonumber \\     
   &~& \hspace{-3.1cm} \left. +16 i C_{Ae}^{'} C_{At}^{'} M_{Z}^2 s \bar{\epsilon
   }^{\overline{k}p_2p_3p_4} s_{W}^2+40 i C_{At}^{'} C_{Ve}^{'} M_{Z}^2 s \bar{\epsilon
   }^{\overline{k}p_2p_3p_4} s_{W}^2+24 i C_{Ve}^{'} C_{Vt}^{'} M_{Z}^2 s \bar{\epsilon
   }^{\overline{k}p_2p_3p_4} s_{W}^2\right. \nonumber \\     
   &~& \hspace{-3.1cm} \left. -16 i C_{Ae}^{'} C_{At}^{'} M_{Z'}^2 s \bar{\epsilon
   }^{\overline{k}p_2p_3p_4} s_{W}^2-40 i C_{At}^{'} C_{Ve}^{'} M_{Z'}^2 s \bar{\epsilon
   }^{\overline{k}p_2p_3p_4} s_{W}^2-24 i C_{Ve}^{'} C_{Vt}^{'} M_{Z'}^2 s \bar{\epsilon
   }^{\overline{k}p_2p_3p_4} s_{W}^2 \right. \nonumber \\ \nonumber
   \end{eqnarray}
   \begin{eqnarray}
   &~& \left. +20 C_{Ae}^{'} C_{At}^{'} M_{Z}^2 s^2 p_1\cdot
   p_3 s_{W}^2+20 C_{At}^{'} C_{Ve}^{'} M_{Z}^2 s^2 p_1\cdot p_3 s_{W}^2+20
   C_{Ae}^{'} C_{Vt}^{'} M_{Z}^2 s^2 p_1\cdot p_3 s_{W}^2 \right. \nonumber \\
   &~& \left. +20 C_{Ve}^{'} C_{Vt}^{'} M_{Z}^2 s^2
   p_1\cdot p_3 s_{W}^2+20 C_{Ae}^{'} C_{At}^{'} M_{Z'}^2 s^2 p_1\cdot
   p_3 s_{W}^2+20 C_{At}^{'} C_{Ve}^{'} M_{Z'}^2 s^2 p_1\cdot p_3 s_{W}^2  \right. \nonumber \\
   &~& \left. +20
   C_{Ae}^{'} C_{Vt}^{'} M_{Z'}^2 s^2 p_1\cdot p_3 s_{W}^2+20 C_{Ve}^{'} C_{Vt}^{'} M_{Z'}^2 s^2
   p_1\cdot p_3 s_{W}^2-20 C_{Ae}^{'} C_{At}^{'} M_{Z}^2 s^2 \left(p_1\cdot
   p_4\right) s_{W}^2 \right. \nonumber \\
   &~& \left. +4 C_{At}^{'} C_{Ve}^{'} M_{Z}^2 s^2 p_1\cdot p_4 s_{W}^2-4
   C_{Ae}^{'} C_{Vt}^{'} M_{Z}^2 s^2 p_1\cdot p_4 s_{W}^2+20 C_{Ve}^{'} C_{Vt}^{'} M_{Z}^2 s^2
   p_1\cdot p_4 s_{W}^2 \right. \nonumber \\
   &~& \left. -20 C_{Ae}^{'} C_{At}^{'} M_{Z'}^2 s^2 p_1\cdot
   p_4 s_{W}^2+4 C_{At}^{'} C_{Ve}^{'} M_{Z'}^2 s^2 p_1\cdot p_4 s_{W}^2-4
   C_{Ae}^{'} C_{Vt}^{'} M_{Z'}^2 s^2 p_1\cdot p_4 s_{W}^2  \right. \nonumber \\
   &~& \left. +20 C_{Ve}^{'} C_{Vt}^{'} M_{Z'}^2 s^2
   p_1\cdot p_4 s_{W}^2-20 C_{Ae}^{'} C_{At}^{'} M_{Z}^2 s^2 p_2\cdot
   p_3 s_{W}^2+4 C_{At}^{'} C_{Ve}^{'} M_{Z}^2 s^2 p_2\cdot p_3 s_{W}^2  \right. \nonumber \\
   &~& \left. -4
   C_{Ae}^{'} C_{Vt}^{'} M_{Z}^2 s^2 p_2\cdot p_3 s_{W}^2 +20 C_{Ve}^{'} C_{Vt}^{'} M_{Z}^2 s^2
   p_2\cdot p_3 s_{W}^2-20 C_{Ae}^{'} C_{At}^{'} M_{Z'}^2 s^2 p_2\cdot
   p_3 s_{W}^2 \right. \nonumber \\
   &~& \left. +4 C_{At}^{'} C_{Ve}^{'} M_{Z'}^2 s^2 p_2\cdot p_3 s_{W}^2-4
   C_{Ae}^{'} C_{Vt}^{'} M_{Z'}^2 s^2 p_2\cdot p_3 s_{W}^2+20 C_{Ve}^{'} C_{Vt}^{'} M_{Z'}^2 s^2
   p_2\cdot p_3 s_{W}^2 \right. \nonumber \\
   &~& \left. -160 C_{Ae}^{'} C_{At}^{'} M_{Z}^2 M_{Z'}^2 p_1\cdot
   p_4 p_2\cdot p_3 s_{W}^2-160 C_{At}^{'} C_{Ve}^{'} M_{Z}^2 M_{Z'}^2
   p_1\cdot p_4 p_2\cdot p_3 s_{W}^2 \right. \nonumber \\
   &~& \left. -160 C_{Ae}^{'} C_{Vt}^{'}
   M_{Z}^2 M_{Z'}^2 p_1\cdot p_4 p_2\cdot p_3 s_{W}^2-160
   C_{Ve}^{'} C_{Vt}^{'} M_{Z}^2 M_{Z'}^2 p_1\cdot p_4 p_2\cdot
   p_3 s_{W}^2 \right. \nonumber \\
   &~& \left. +80 C_{Ae}^{'} C_{At}^{'} M_{Z}^2 s p_1\cdot p_4
   p_2\cdot p_3 s_{W}^2+32 C_{At}^{'} C_{Ve}^{'} M_{Z}^2 s p_1\cdot
   p_4 p_2\cdot p_3 s_{W}^2+48 C_{Ae}^{'} C_{Vt}^{'} M_{Z}^2 s
   p_1\cdot p_4 p_2\cdot p_3 s_{W}^2 \right. \nonumber \\
   &~& \left. +80 C_{Ae}^{'} C_{At}^{'}
   M_{Z'}^2 s p_1\cdot p_4 p_2\cdot p_3 s_{W}^2+32 C_{At}^{'}
   C_{Ve}^{'} M_{Z'}^2 s p_1\cdot p_4 p_2\cdot p_3 s_{W}^2+48
   C_{Ae}^{'} C_{Vt}^{'} M_{Z'}^2 s p_1\cdot p_4 p_2\cdot p_3
   s_{W}^2 \right. \nonumber \\
   &~& \left. +20 C_{Ae}^{'} C_{At}^{'} M_{Z}^2 s^2 p_2\cdot p_4 s_{W}^2+20 C_{At}^{'} C_{Ve}^{'}
   M_{Z}^2 s^2 p_2\cdot p_4 s_{W}^2+20 C_{Ae}^{'} C_{Vt}^{'} M_{Z}^2 s^2 p_2\cdot
   p_4 s_{W}^2 \right. \nonumber \\
   &~& \left.  +20 C_{Ve}^{'} C_{Vt}^{'} M_{Z}^2 s^2 p_2\cdot p_4 s_{W}^2+20
   C_{Ae}^{'} C_{At}^{'} M_{Z'}^2 s^2 p_2\cdot p_4 s_{W}^2+20 C_{At}^{'} C_{Ve}^{'} M_{Z'}^2 s^2
   p_2\cdot p_4 s_{W}^2 \right. \nonumber \\
   &~& \left.  +20 C_{Ae}^{'} C_{Vt}^{'} M_{Z'}^2 s^2 p_2\cdot
   p_4 s_{W}^2+20 C_{Ve}^{'} C_{Vt}^{'} M_{Z'}^2 s^2 p_2\cdot p_4 s_{W}^2+160
   C_{Ae}^{'} C_{At}^{'} M_{Z}^2 M_{Z'}^2 p_1\cdot p_3 p_2\cdot
   p_4 s_{W}^2 \right. \nonumber \\
   &~& \left. -32 C_{At}^{'} C_{Ve}^{'} M_{Z}^2 M_{Z'}^2 p_1\cdot p_3
   p_2\cdot p_4 s_{W}^2+32 C_{Ae}^{'} C_{Vt}^{'} M_{Z}^2 M_{Z'}^2 p_1\cdot
   p_3 p_2\cdot p_4 s_{W}^2 \right. \nonumber \\
   &~& \left. -160 C_{Ve}^{'} C_{Vt}^{'} M_{Z}^2 M_{Z'}^2
   p_1\cdot p_3 p_2\cdot p_4 s_{W}^2-80 C_{Ae}^{'} C_{At}^{'}
   M_{Z}^2 s p_1\cdot p_3 p_2\cdot p_4 s_{W}^2-32 C_{At}^{'}
   C_{Ve}^{'} M_{Z}^2 s p_1\cdot p_3 p_2\cdot p_4 s_{W}^2 \right. \nonumber \\
   &~& \left. -48
   C_{Ae}^{'} C_{Vt}^{'} M_{Z}^2 s p_1\cdot p_3 p_2\cdot p_4
   s_{W}^2-80 C_{Ae}^{'} C_{At}^{'} M_{Z'}^2 s p_1\cdot p_3 p_2\cdot
   p_4 s_{W}^2-32 C_{At}^{'} C_{Ve}^{'} M_{Z'}^2 s p_1\cdot p_3
   p_2\cdot p_4 s_{W}^2 \right. \nonumber \\
   &~& \left. -48 C_{Ae}^{'} C_{Vt}^{'} M_{Z'}^2 s p_1\cdot
   p_3 p_2\cdot p_4 s_{W}^2+3 C_{Ae}^{'} C_{At}^{'} M_{Z}^2 s^3+3 C_{At}^{'}
   C_{Ve}^{'} M_{Z}^2 s^3+3 C_{Ae}^{'} C_{Vt}^{'} M_{Z}^2 s^3 \right. \nonumber \\
   &~& \left. +3 C_{Ve}^{'} C_{Vt}^{'} M_{Z}^2 s^3+3 C_{Ae}^{'} C_{At}^{'} M_{Z'}^2 s^3+3
   C_{At}^{'} C_{Ve}^{'} M_{Z'}^2 s^3+3 C_{Ae}^{'} C_{Vt}^{'} M_{Z'}^2 s^3+3 C_{Ve}^{'} C_{Vt}^{'} M_{Z'}^2 s^3 \right. \nonumber \\
   &~& \left. -12 C_{Ae}^{'} C_{At}^{'}
   m_t^2 M_{Z}^2 M_{Z'}^2 s-12 C_{At}^{'} C_{Ve}^{'} m_t^2 M_{Z}^2 M_{Z'}^2 s+12 C_{Ae}^{'} C_{Vt}^{'} m_t^2 M_{Z}^2
   M_{Z'}^2 s+12 C_{Ve}^{'} C_{Vt}^{'} m_t^2 M_{Z}^2 M_{Z'}^2 s \right. \nonumber \\
   &~& \left. +6 i C_{Ae}^{'} C_{At}^{'} M_{Z}^2 s \bar{\epsilon
   }^{\overline{k}p_1p_2p_3}+6 i C_{At}^{'} C_{Ve}^{'} M_{Z}^2 s \bar{\epsilon
   }^{\overline{k}p_1p_2p_3}+6 i C_{Ae}^{'} C_{Vt}^{'} M_{Z}^2 s \bar{\epsilon
   }^{\overline{k}p_1p_2p_3}+6 i C_{Ve}^{'} C_{Vt}^{'} M_{Z}^2 s \bar{\epsilon
   }^{\overline{k}p_1p_2p_3} \right. \nonumber \\
   &~& \left. -6 i C_{Ae}^{'} C_{At}^{'} M_{Z'}^2 s \bar{\epsilon
   }^{\overline{k}p_1p_2p_3}-6 i C_{At}^{'} C_{Ve}^{'} M_{Z'}^2 s \bar{\epsilon
   }^{\overline{k}p_1p_2p_3}-6 i C_{Ae}^{'} C_{Vt}^{'} M_{Z'}^2 s \bar{\epsilon
   }^{\overline{k}p_1p_2p_3}-6 i C_{Ve}^{'} C_{Vt}^{'} M_{Z'}^2 s \bar{\epsilon
   }^{\overline{k}p_1p_2p_3} \right. \nonumber \\
   &~& \left. +6 i C_{Ae}^{'} C_{At}^{'} M_{Z}^2 s \bar{\epsilon
   }^{\overline{k}p_1p_2p_4}+6 i C_{At}^{'} C_{Ve}^{'} M_{Z}^2 s \bar{\epsilon
   }^{\overline{k}p_1p_2p_4}+6 i C_{Ae}^{'} C_{Vt}^{'} M_{Z}^2 s \bar{\epsilon
   }^{\overline{k}p_1p_2p_4}+6 i C_{Ve}^{'} C_{Vt}^{'} M_{Z}^2 s \bar{\epsilon
   }^{\overline{k}p_1p_2p_4} \right. \nonumber \\
   &~& \left. -6 i C_{Ae}^{'} C_{At}^{'} M_{Z'}^2 s \bar{\epsilon
   }^{\overline{k}p_1p_2p_4}-6 i C_{At}^{'} C_{Ve}^{'} M_{Z'}^2 s \bar{\epsilon
   }^{\overline{k}p_1p_2p_4}-6 i C_{Ae}^{'} C_{Vt}^{'} M_{Z'}^2 s \bar{\epsilon
   }^{\overline{k}p_1p_2p_4}-6 i C_{Ve}^{'} C_{Vt}^{'} M_{Z'}^2 s \bar{\epsilon
   }^{\overline{k}p_1p_2p_4} \right. \nonumber \\
   &~& \left. -6 i C_{Ae}^{'} C_{At}^{'} M_{Z}^2 s \bar{\epsilon
   }^{\overline{k}p_1p_3p_4}-6 i C_{At}^{'} C_{Ve}^{'} M_{Z}^2 s \bar{\epsilon
   }^{\overline{k}p_1p_3p_4}-6 i C_{Ae}^{'} C_{Vt}^{'} M_{Z}^2 s \bar{\epsilon
   }^{\overline{k}p_1p_3p_4}-6 i C_{Ve}^{'} C_{Vt}^{'} M_{Z}^2 s \bar{\epsilon
   }^{\overline{k}p_1p_3p_4} \right. \nonumber \\
   &~& \left. +6 i C_{Ae}^{'} C_{At}^{'} M_{Z'}^2 s \bar{\epsilon
   }^{\overline{k}p_1p_3p_4}+6 i C_{At}^{'} C_{Ve}^{'} M_{Z'}^2 s \bar{\epsilon
   }^{\overline{k}p_1p_3p_4}+6 i C_{Ae}^{'} C_{Vt}^{'} M_{Z'}^2 s \bar{\epsilon
   }^{\overline{k}p_1p_3p_4}+6 i C_{Ve}^{'} C_{Vt}^{'} M_{Z'}^2 s \bar{\epsilon
   }^{\overline{k}p_1p_3p_4} \right. \nonumber \\
   &~& \left. -6 i C_{Ae}^{'} C_{At}^{'} M_{Z}^2 s \bar{\epsilon
   }^{\overline{k}p_2p_3p_4}-6 i C_{At}^{'} C_{Ve}^{'} M_{Z}^2 s \bar{\epsilon
   }^{\overline{k}p_2p_3p_4}-6 i C_{Ae}^{'} C_{Vt}^{'} M_{Z}^2 s \bar{\epsilon
   }^{\overline{k}p_2p_3p_4}-6 i C_{Ve}^{'} C_{Vt}^{'} M_{Z}^2 s \bar{\epsilon
   }^{\overline{k}p_2p_3p_4} \right. \nonumber
   \end{eqnarray}
   \begin{eqnarray}
   &~& \left. +6 i C_{Ae}^{'} C_{At}^{'} M_{Z'}^2 s \bar{\epsilon
   }^{\overline{k}p_2p_3p_4}+6 i C_{At}^{'} C_{Ve}^{'} M_{Z'}^2 s \bar{\epsilon
   }^{\overline{k}p_2p_3p_4}+6 i C_{Ae}^{'} C_{Vt}^{'} M_{Z'}^2 s \bar{\epsilon
   }^{\overline{k}p_2p_3p_4}+6 i C_{Ve}^{'} C_{Vt}^{'} M_{Z'}^2 s \bar{\epsilon
   }^{\overline{k}p_2p_3p_4} \right. \nonumber\\
   &~& \left. -6 C_{Ae}^{'} C_{At}^{'} M_{Z}^2 s^2 \left(p_1\cdot
   p_3\right)-6 C_{At}^{'} C_{Ve}^{'} M_{Z}^2 s^2 p_1\cdot p_3-6 C_{Ae}^{'} C_{Vt}^{'}
   M_{Z}^2 s^2 p_1\cdot p_3-6 C_{Ve}^{'} C_{Vt}^{'} M_{Z}^2 s^2 p_1\cdot p_3 \right. \nonumber\\
   &~& \left. -6 C_{Ae}^{'} C_{At}^{'} M_{Z'}^2 s^2 p_1\cdot p_3-6 C_{At}^{'} C_{Ve}^{'}
   M_{Z'}^2 s^2 p_1\cdot p_3-6 C_{Ae}^{'} C_{Vt}^{'} M_{Z'}^2 s^2 p_1\cdot p_3 -6 C_{Ve}^{'} C_{Vt}^{'} M_{Z'}^2 s^2 p_1\cdot p_3 \right. \nonumber\\
   &~& \left. +48 C_{Ae}^{'} C_{At}^{'}
   M_{Z}^2 M_{Z'}^2 p_1\cdot p_4 p_2\cdot p_3+48 C_{At}^{'}
   C_{Ve}^{'} M_{Z}^2 M_{Z'}^2 p_1\cdot p_4 p_2\cdot p_3+48
   C_{Ae}^{'} C_{Vt}^{'} M_{Z}^2 M_{Z'}^2 p_1\cdot p_4 p_2\cdot
   p_3  \right. \nonumber\\
   &~& \left. +48 C_{Ve}^{'} C_{Vt}^{'} M_{Z}^2 M_{Z'}^2 p_1\cdot p_4
   p_2\cdot p_3-12 C_{Ae}^{'} C_{At}^{'} M_{Z}^2 s p_1\cdot p_4
   p_2\cdot p_3-12 C_{At}^{'} C_{Ve}^{'} M_{Z}^2 s p_1\cdot p_4
   p_2\cdot p_3 \right. \nonumber\\
   &~& \left.  -12 C_{Ae}^{'} C_{Vt}^{'} M_{Z}^2 s p_1\cdot p_4
   p_2\cdot p_3-12 C_{Ve}^{'} C_{Vt}^{'} M_{Z}^2 s p_1\cdot p_4
   p_2\cdot p_3-12 C_{Ae}^{'} C_{At}^{'} M_{Z'}^2 s p_1\cdot p_4
   p_2\cdot p_3  \right. \nonumber\\
   &~& \left. -12 C_{At}^{'} C_{Ve}^{'} M_{Z'}^2 s p_1\cdot p_4
   p_2\cdot p_3-12 C_{Ae}^{'} C_{Vt}^{'} M_{Z'}^2 s p_1\cdot p_4
   p_2\cdot p_3-12 C_{Ve}^{'} C_{Vt}^{'} M_{Z'}^2 s p_1\cdot p_4
   p_2\cdot p_3 \right. \nonumber\\
   &~& \left. -6 C_{Ae}^{'} C_{At}^{'} M_{Z}^2 s^2 p_2\cdot p_4-6
   C_{At}^{'} C_{Ve}^{'} M_{Z}^2 s^2 p_2\cdot p_4-6 C_{Ae}^{'} C_{Vt}^{'} M_{Z}^2 s^2
   p_2\cdot p_4-6 C_{Ve}^{'} C_{Vt}^{'} M_{Z}^2 s^2 p_2\cdot p_4 \right. \nonumber\\
   &~& \left. -6
   C_{Ae}^{'} C_{At}^{'} M_{Z'}^2 s^2 p_2\cdot p_4-6 C_{At}^{'} C_{Ve}^{'} M_{Z'}^2 s^2
   p_2\cdot p_4-6 C_{Ae}^{'} C_{Vt}^{'} M_{Z'}^2 s^2 p_2\cdot
   p_4 -6 C_{Ve}^{'} C_{Vt}^{'} M_{Z'}^2 s^2 p_2\cdot p_4 \right. \nonumber\\
   &~& \left. +12 C_{Ae}^{'} C_{At}^{'}
   M_{Z}^2 s p_1\cdot p_3 p_2\cdot p_4+12 C_{At}^{'} C_{Ve}^{'}
   M_{Z}^2 s p_1\cdot p_3 p_2\cdot p_4+12 C_{Ae}^{'} C_{Vt}^{'}
   M_{Z}^2 s p_1\cdot p_3 p_2\cdot p_4 \right. \nonumber\\
   &~& \left. +12 C_{Ve}^{'} C_{Vt}^{'}
   M_{Z}^2 s p_1\cdot p_3 p_2\cdot p_4+12 C_{Ae}^{'} C_{At}^{'}
   M_{Z'}^2 s p_1\cdot p_3 p_2\cdot p_4+12 C_{At}^{'} C_{Ve}^{'}
   M_{Z'}^2 s p_1\cdot p_3 p_2\cdot p_4 \right. \nonumber\\
   &~& \left. +12 C_{Ae}^{'} C_{Vt}^{'}
   M_{Z'}^2 s p_1\cdot p_3 p_2\cdot p_4+12 C_{Ve}^{'} C_{Vt}^{'}
   M_{Z'}^2 s p_1\cdot p_3 p_2\cdot p_4\right)
   \label{MZZp-four-formula}
 \end{eqnarray}
 \end{appendices}
%%%%%%%%%%%%%%%%%%%%%%%%%%%%%%%%%%%%%%%%%%%%%%%%%%%%%%%%%%%%%%%   

 \end{document}